\documentclass[acmsmall,screen,nonacm]{acmart}

\AtBeginDocument{%
  }

\setcopyright{none}
\usepackage{longtable} 
\usepackage{multirow}
\usepackage{array}
\usepackage[capitalize]{cleveref}

\usepackage{tcolorbox}
\newtcolorbox{findingbox}[1]{
  colframe=cyan!80!blue,     
  colback=cyan!5!white,             
  coltitle=white,             
  fonttitle=\bfseries,
  title={#1},
  arc=4pt,
  boxrule=0.5pt,
  left=5pt, right=5pt, top=5pt, bottom=5pt
}

\begin{document}

\title{From Collaboration to Regulation: Characterizing Governance Practice in Three Deep Learning Open Source Communities}

\author{Ruiqiao Qiu}
\orcid{0009-0002-2313-4581}
\email{qrq2001@stu.pku.edu.cn}
\affiliation{%
  \department{School of Computer Science}
  \institution{Peking University}
  \city{Beijing}
  \country{China}
}
\affiliation{%
  \institution{Key Laboratory of High Confidence Software Technologies, Ministry of Education}
  \city{Beijing}
  \country{China}
}

\author{Wenhao Yang}
\email{yangwh@stu.pku.edu.cn}
\affiliation{%
  \department{School of Computer Science}
  \institution{Peking University}
  \city{Beijing}
  \country{China}
}
\affiliation{%
  \institution{Key Laboratory of High Confidence Software Technologies, Ministry of Education}
  \city{Beijing}
  \country{China}
}

\author{Minghui Zhou}
\orcid{0000-0001-6324-3964}
\email{zhmh@pku.edu.cn}
\authornote{Corresponding author.}
\affiliation{%
  \department{School of Computer Science}
  \institution{Peking University}
  \city{Beijing}
  \country{China}
}
\affiliation{%
  \institution{Key Laboratory of High Confidence Software Technologies, Ministry of Education}
  \city{Beijing}
  \country{China}
}

\renewcommand{\shortauthors}{Qiu et al.}

\begin{abstract}
    Collaboration in Open Source Software (OSS) projects involves substantial coordination and quality-control challenges across a diverse contributor base. For contributors, these challenges often appear as unclear participation pathways and repeated clarification during contribution and review. 
    Projects address such problems through documented governance rules that structure participation and collaboration. Yet maintainers have limited systematic guidance on what rules to codify, when to introduce or revise them, and how to organize them across community documents.
    To bridge this gap, we conducted a mixed-methods empirical study on three mature deep learning (DL) frameworks: PyTorch, TensorFlow, and Paddle. 
    Adopting the Institutional Analysis and Development (IAD) framework from institutional theory, we contextualized seven types of governance rules within the OSS context, such as community roles (position rules), contribution guidelines (choice rules), and contribution standards (scope rules).
    We collected 109 documents from the three DL frameworks and employed thematic analysis to identify how governance rules are represented in documentation, resulting in 17 rule themes categorized into the seven rule types. 
    We found that rule themes were distributed across document categories in functionally distinct patterns. While operational rules (e.g., workflows) appear across all three projects, structural rules (e.g., hierarchies) vary substantially across projects. 
    By tracing the history of these documents through over 1,700 commits, we reconstructed when rule themes first appeared in the collected documentation, how they were revised, and how their rule-bearing text accumulated across documentation files. Operational rule themes generally appeared earlier in the collected documentation and were revised more frequently, whereas many structural rule themes appeared later and changed less often thereafter. We also observed increasing documentation specialization, as rule-bearing content accumulated across task-specific or role-specific files over project history.  
    We further identified four governance functions reflected in substantive rule changes: Norm Alignment, Workflow Refinement, Coordination Structuring, and Community and Governance Development. 
    Synthesizing these findings, we derive 33 actionable governance practices that connect rule content, longitudinal patterns of public codification and revision, and the governance functions reflected in rule changes. These practices are most directly relevant to mature, large-scale deep learning OSS projects with substantial coordination demands and organizational involvement, while also providing reference points for projects facing similar governance conditions.
\end{abstract}




\maketitle

\section{Introduction}
Software development is a collective activity that relies heavily on collaboration among developers. However, collaboration is challenging because communication, coordination, and consensus-building can be time-consuming~\cite{brooks1974mythical}. In Open Source Software (OSS) projects, this challenge is amplified not only by scale but also by the nature of interaction: developers around the world contribute through open online communities, and collaboration is fundamentally asynchronous and artifact-mediated~\cite{GitHubOctoverse2025}.  
Unlike traditional teams where implicit norms can be conveyed interpersonally, OSS communities rely on explicit knowledge encoded in documentation to define basic operations, such as how to configure environments, distinguish roles, and follow contribution workflows. Without these encoded rules, potential contributors lack the entry points to participate. 
Furthermore, the complexity of these interactions is compounded by the diversity of global contributors. Variations in expertise~\cite{bock2023automatic,lee2017understanding,li2024preliminary}, motivation~\cite{zhang2024paid}, background~\cite{shameer2023relationship,wang2025uncovering,bhuiyan2025developers,frluckaj2025diverse,dutta2023diversity}, and personal characteristics~\cite{sarker2025landscape,bharadwaj2025shifting,weichbroth2025survey} often result in misaligned behaviors. The growing commercial involvement in OSS projects adds another layer of complexity~\cite{zhou2016inflow, zhang2022commercial, zhang2022corporate,qin2025developers}, as corporate interests may diverge from community values. Consequently, the absence of clear governance rules not only creates barriers to entry but also hinders tasks like contribution evaluation and code review~\cite{steinmacher2018almost,li2020redundancy, gousios2015work}, which may significantly hinder development progress and place a heavy burden on maintainers~\cite{linaaker2024sustaining,russo2024shock}.  

To manage these complexities, OSS communities commonly adopt documented rules to regulate participation, contribution, communication, and decision-making. These rules can take many forms, including contributing guidelines, codes of conduct, governance structure files, and pull request templates. 
While prior studies often refer to specific subsets of these documents with terms such as \textit{onboarding guidelines} or \textit{contribution rules}, these terms typically focus on particular entry points or narrow processes. In this paper, we adopt the broader term \textbf{governance rules} to capture the full range of documented prescriptions that structure how collaboration is coordinated and controlled—from contribution workflows and role assignments to behavioral norms and decision-making procedures. This usage aligns with prior work in organizational theory and open source research that treats governance as a multi-dimensional construct spanning coordination, control, and community maintenance~\cite{markus2007governance,de2011governance}. 

However, formulating effective governance is far from trivial. 
In practice, maintainers frequently introduce or revise rules only after a collaboration problem has already emerged~\cite{gaughan2025introduction}, rather than anticipating needs in advance. Consequently, many projects suffer from fragmented or missing guidelines. This reveals that many maintainers lack a systematic understanding of what rules are necessary, when rules should be introduced during a project's lifecycle, and why certain regulations are crucial for specific collaboration contexts. This blind spot may leave projects less prepared to handle recurring coordination problems, highlighting the need for empirically grounded guidance on how OSS communities formulate and maintain documented governance rules. Gaining a deeper understanding of documented governance rules can help maintainers reason more systematically about collaboration support in OSS projects.

In this paper, we aim to develop a structured framework for examining the codification, timing, and evolution of governance rules in OSS communities, and to derive actionable practices for maintainers.
In particular, we formulate three research questions:

First, while existing literature acknowledges the importance of governance, we still lack an integrated understanding of how governance is actually instantiated in documentation. Rules are often scattered across various files (e.g., \texttt{CONTRIBUTING.md}, \texttt{README.md}, issue templates), making it difficult to understand how governance is distributed across documentation. Furthermore, it remains unclear whether these governance structures are commonly shared or if they vary significantly across different projects.
To systematize this knowledge, we ask:

\textbf{RQ1: What specific governance rules are codified in OSS projects, how are they structured across documentation, and how do they vary across communities?} 

Second, governance is not static; rules are revised and reorganized as project conditions change. However, little is known about when specific rule themes first appear in project documentation, how they are modified, and how rule-bearing content becomes distributed across files as coordination needs grow. Understanding these temporal dynamics can help maintainers reason more systematically about governance maintenance. Therefore, we investigate:

\textbf{RQ2: How do governance rules evolve throughout the project lifecycle in terms of timing, frequency, and documentation structure?}

Finally, identifying when governance rules are introduced or modified does not fully explain what role these changes play in project governance. Although rule-bearing text is repeatedly revised throughout a project's lifecycle, prior work provides limited understanding of the governance functions reflected in these modifications. Examining these functions can clarify how documented rule changes support the organization and maintenance of OSS collaboration. Therefore, we ask:

\textbf{RQ3: What governance functions are reflected in the formulation and modification of governance rules?}

we operationalized Ostrom's Institutional Analysis and Development (IAD) framework to classify OSS governance rules using its seven rule types.

We then conducted an empirical study on three mature deep learning OSS projects—PyTorch, TensorFlow, and Paddle. Deep learning frameworks provide a suitable empirical setting because they are widely used, rapidly evolving, and technically complex, which creates substantial demands for contribution management, review coordination, community communication, and newcomer onboarding. Focusing on projects from the same technical domain improves cross-project comparability, while their publicly accessible governance documents make it possible to systematically trace both the content and evolution of governance rules.

For RQ1, we performed a thematic analysis of the latest versions of community documents and mapped the identified rule-bearing segments to the IAD framework. We identified 17 rule themes mapped to seven IAD rule types, such as position, choice, information, and scope rules. We further examined how these themes are distributed across document categories. The resulting distribution shows that operational themes, such as contribution submission guidance and code acceptance criteria, are frequently codified in contributing files and templates, whereas structural themes, such as role definitions and role acquisition processes, are mainly documented in governance structure files. At the project level, operational themes are broadly shared across the three projects, while structural and incentive-related themes show more visible variation.

For RQ2, we reconstructed the evolutionary history of governance rule themes by mining the version histories of their corresponding documents. We traced over 1,700 commits, handled file renames and history breaks, and identified substantive rule-modifying records for analyzing when rule themes first appeared in documentation and how frequently they were modified. We also recorded which logical files contained substantive rule-bearing text associated with each rule theme to examine documentation splitting over time. Operational rule themes generally appeared earlier in the collected documentation and were modified more frequently, whereas many structural rule themes appeared later and changed less often thereafter. Additionally, we observed increasing documentation specialization, where rule-bearing content accumulated across task-specific or role-specific files, such as workflow-specific contribution guides and SIG-related documents.

For RQ3, we analyzed the substantive rule-modifying records identified in RQ2. We examined the semantic diff, the affected rule theme, and the surrounding document context of each record to identify the primary governance function reflected in the modified rule-bearing text. Through an iterative coding process, we identified four recurring governance functions: Norm Alignment, Workflow Refinement, Coordination Structuring, and Community and Governance Development.

Based on the integrated findings from RQ1 through RQ3, we synthesized 33 actionable governance practices for OSS maintainers. The practices are organized around the 17 rule themes identified in RQ1. The longitudinal findings from RQ2 inform the conditions under which similar rules first appeared in documentation, were refined, or were reorganized, while the four governance functions identified in RQ3 clarify how these practices support norm alignment, workflow refinement, coordination structuring, and community and governance development. Together, the practices provide a structured reference for formulating and maintaining documented governance rules under different coordination and organizational conditions.

By integrating an IAD-based taxonomy, longitudinal document analysis, and thematic analysis of governance functions, this study provides an empirically grounded account of how governance rules are codified, organized, and revised in mature OSS projects. In general, the main contributions of this paper are:

\begin{itemize}
    \item An IAD-informed taxonomy for classifying documentary governance rules in OSS projects.
    \item A longitudinal analysis characterizing when rule themes first appeared in documentation, how frequently they were modified, and how their documentation specialized over time.
    \item A thematic analysis identifying four governance functions reflected in the formulation and modification of documented governance rules.
    \item A set of 33 actionable governance practices synthesized from the rule themes, longitudinal patterns, and governance functions identified in the study.
\end{itemize}

\section{Background and Related Work}

\subsection{Governance as OSS Collaboration Mechanism} ~\label{ss:governance-as-collaboration-mechanisms}

As highlighted in the introduction, the open and distributed nature of OSS development introduces substantial coordination overhead. To reduce uncertainty and coordinate distributed work, OSS communities rely on community documents such as \texttt{README.md}, \texttt{CONTRIBUTING.md}, codes of conduct, issue templates, pull-request templates, and \texttt{GOVERNANCE.md} files. Prior work has often examined these documents through the lenses of knowledge transfer, onboarding, and contribution support. For example, README and contributing files support knowledge transfer~\cite{prana2019categorizing,okong2025knowledge} and signal a project's friendliness to newcomers~\cite{wang2023study}. Clear contributing guidelines lower onboarding barriers by clarifying the contribution flow~\cite{fronchetti2023contributing}, while explicit guidelines on development workflows, coding styles, and testing protocols~\cite{kruger2020can,turzo2025first} help contributors prepare changes that meet project expectations before review.

\begin{figure}[h]
  \centering
  \includegraphics[width=0.8\linewidth]{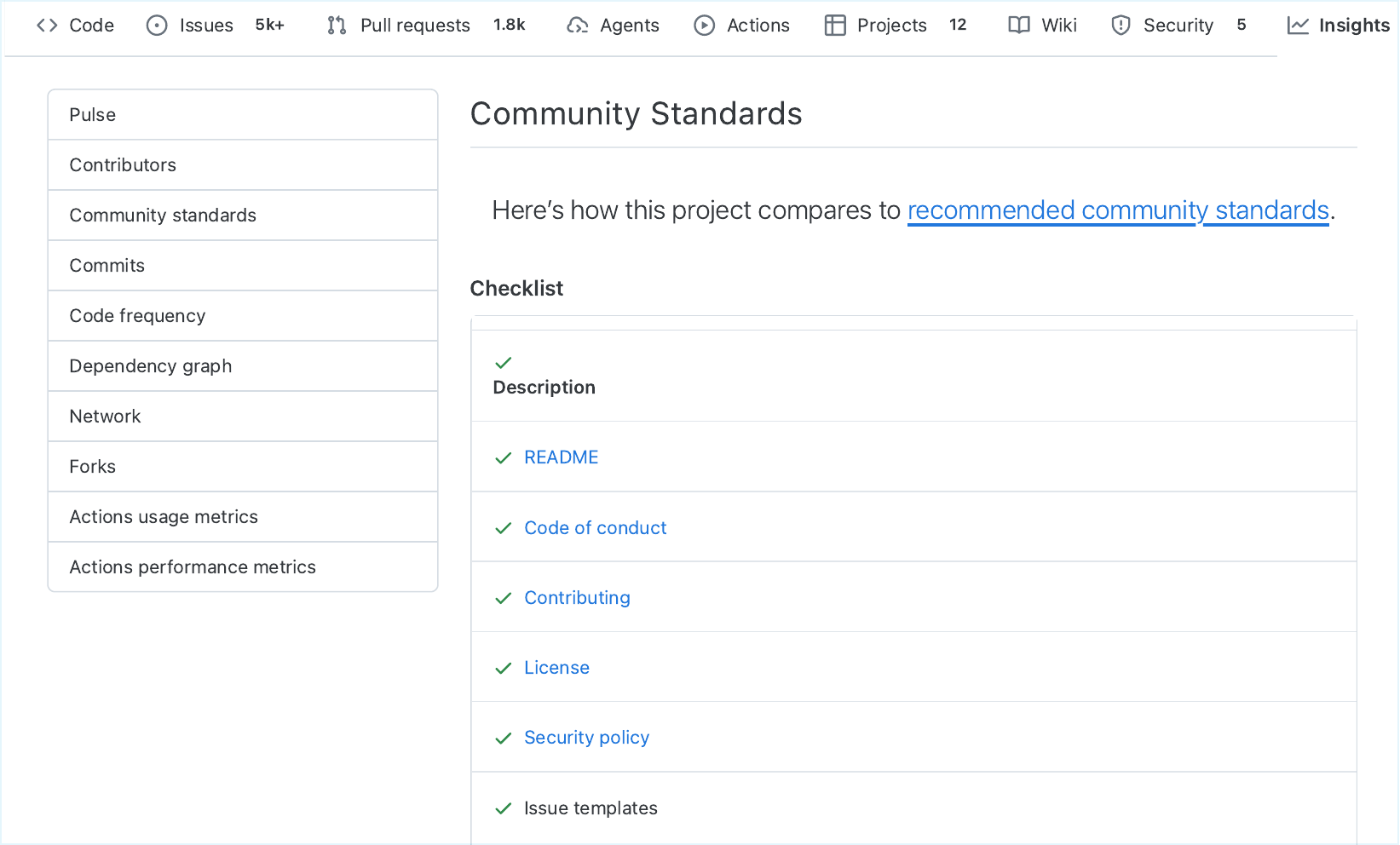}
  \caption{Community Standards Checklist on GitHub}
  \label{fig:community_standards}
\end{figure}

The centrality of these documents is further reflected in the infrastructure provided by collaborative development platforms. GitHub integrates a "Community Standards" checklist in its project insights page, as shown in \cref{fig:community_standards}, explicitly monitoring the presence of important artifacts such as codes of conduct, contribution guidelines, and issue templates. This checklist highlights the platform-level importance of formalized community documentation. However, while such checklists monitor the existence of these files, they provide little guidance on their substance, leaving maintainers to determine the specific rules and workflows themselves.

In this study, we examine these documents from a governance perspective. Prior OSS governance research uses governance to describe how communities allocate roles, responsibilities, authority, decision processes, and control mechanisms~\cite{markus2007governance,chulani2008software,jensen2010governance,o2007governance}. This perspective allows us to connect contribution-facing documents with broader community regulation. For example, onboarding and contribution guidelines do not only help newcomers submit patches; they also specify who can participate, what actions contributors should take, what information should be provided, and what outcomes are acceptable. Other artifacts, such as codes of conduct, issue and pull-request templates, and \texttt{GOVERNANCE.md} files, further regulate communication norms, role structures, and community responsibilities~\cite{yan2023github}. We therefore use the term \textit{governance rules} to refer to documented prescriptions that structure participation and coordination in OSS communities.

\subsection{Studies on OSS Governance} ~\label{ss:dimensions-of-governance}
OSS governance is widely recognized as a complex, multidimensional construct rather than a singular process. 
Markus et al.~\cite{markus2007governance} conceptualized governance as spanning diverse areas including asset ownership, conflict resolution, and software development processes. Chulani et al.~\cite{chulani2008software} distinguished between the "static structure" of governance (e.g., chains of responsibility, authority rights) and its "dynamic execution" (e.g., control mechanisms and measurements), highlighting the dual nature of governance in both organizational form and operational process. Furthermore, scholars have analyzed governance across different levels of granularity. Jensen and Scacchi~\cite{jensen2010governance} proposed a multi-level framework, examining governance interactions at the micro (individual artifacts and actions), meso (project team collaboration and leadership), and macro (inter-project ecosystem) levels. De Noni et al.~\cite{de2011governance} further validated the interdependence of these dimensions, and found that leadership structures are often intrinsically linked to intellectual property management strategies.

Some researchers have developed various taxonomies to further describe the specific components of governance rules. Izquierdo et al.~\cite{izquierdo2015enabling} classified governance rules into sets of contribution and acceptance protocols. Recently, Yin et al.~\cite{yin2022open} extracted 12 distinct categories of "institutional statements" from developer emails. Building on this semantic approach, Chakraborti et al.~\cite{chakraborti2024we} further investigated "policy internalization" by quantifying the alignment between these developer discussions and the foundation's formal regulations. 

Beyond structural dimensions, O'Mahony et al.~\cite{o2007governance} identified the normative principles that distinguish community-driven projects from those dominated by single commercial entities. They argued that successful mature projects must embody core values such as independence, pluralism, representation, decentralized decision-making, and autonomous participation.

\subsection{Studies on OSS Collaboration Practice} ~\label{ss:collaboration-practices}

Communities have evolved a myriad of specific collaboration practices that function as practical governance mechanisms. Researchers in software engineering have extensively investigated these practices, ranging from technical contribution workflows to social interaction norms.

A primary focus of existing literature is the governance of entry. Since the quality of contributions directly affects the project’s outcomes, communities establish implicit or explicit rules to filter patches. Alami et al.~\cite{alami2021pull} identified governance styles for PRs (e.g., protective) and principles maintainers use to assess PRs. However, strict governance can create barriers for newcomers~\cite{steinmacher2015systematic}. To balance quality control with openness, researchers have examined mechanisms that lower entry barriers, such as the usage of "good first issues" in guiding newcomers~\cite{tan2020first, xiao2022recommending, xiao2023personalized} and mentoring programs like Google Summer of Code~\cite{tan2023understanding}. These studies show that entry governance involves both assessing contributions and designing pathways through which newcomers can identify suitable tasks, receive support, and gradually participate in the community.

Beyond entry, governance mechanisms are essential for regulating the internal workflow and workload distribution. Peer code review mechanisms have been identified not only as quality-assurance tools but also as mechanisms for trust-building and impression formation among peers~\cite{alami2019does, bosu2013impact}. While such mechanisms support collaboration and quality control, studies also reveal their limitations at scale. Zhou et al.~\cite{zhou2017scalability} analyzed the growth of the Linux Kernel and found that maintainers' workload distribution is highly imbalanced across modules, raising concerns about the long-term scalability and sustainability of governance structures. To mitigate such bottlenecks, communities adopt specific role structures and review hierarchies. Tan et al.~\cite{tan2020scaling} observed the shift towards a multiple-committer model in Linux Kernel community to decentralize review authority. These findings suggest that governance rules are important ways to maintain code quality while preventing maintainer burnout.

Effective collaboration also relies on the governance of interaction and behavioral standards. Given the distributed nature of OSS, developers must navigate various synchronous and asynchronous communication channels~\cite{geiger2018types, tenorio2018accountability,ebert2022communication}. Prior work has also examined how contributors frame information when submitting patches~\cite{tan2019communicate}, and how standardized tools like issue templates and PR templates help improve message exchange and reduce friction~\cite{li2022follow, li2023first, sulun2024empirical}. In addition to technical interactions, maintaining a healthy community culture is important. The adoption of Code of Conduct has become a standard governance practice to regulate behavior, promote diversity, and resolve conflicts~\cite{li2021code}.

\subsection{Research Gap} 
Although the role of governance rules as coordination and control mechanisms is well established (\cref{ss:governance-as-collaboration-mechanisms}), practical support for systematically designing such rules remains limited. For example, platforms such as GitHub monitor and encourage the presence of governance artifacts through checklists, yet they offer minimal guidance on their actual content or effectiveness.
In parallel, the academic literature provides limited actionable support for maintainers. As outlined in \cref{ss:dimensions-of-governance} and \cref{ss:collaboration-practices}, a clear disconnect exists between abstract theoretical models of governance and concrete empirical studies of collaborative practices. 
On the one hand, theoretical models offer broad insights into structures and values, but typically lack operational specificity. On the other hand, software engineering studies present targeted tactics for isolated practices, yet do not provide a unified perspective that links these tactics within a coherent governance strategy. Furthermore, both streams tend to treat governance as a static construct, overlooking the need for governance mechanisms to evolve in alignment with project maturity.

As a result, maintainers are often left navigating a fragmented landscape: they may recognize the importance of governance and know about individual tools, but lack a structured methodology for adapting these mechanisms over time.

To fill this gap, we combine the analytical depth of the IAD framework (which will be introduced in \cref{ss:iadframework}) with detailed empirical studies of open-source project trajectories. By mapping documentary governance rules to theoretical categories and tracing their evolution, we aim to develop an empirically grounded framework for examining OSS governance.

\section{Methodology}

\begin{figure}[h]
  \centering
  \includegraphics[width=1\linewidth]{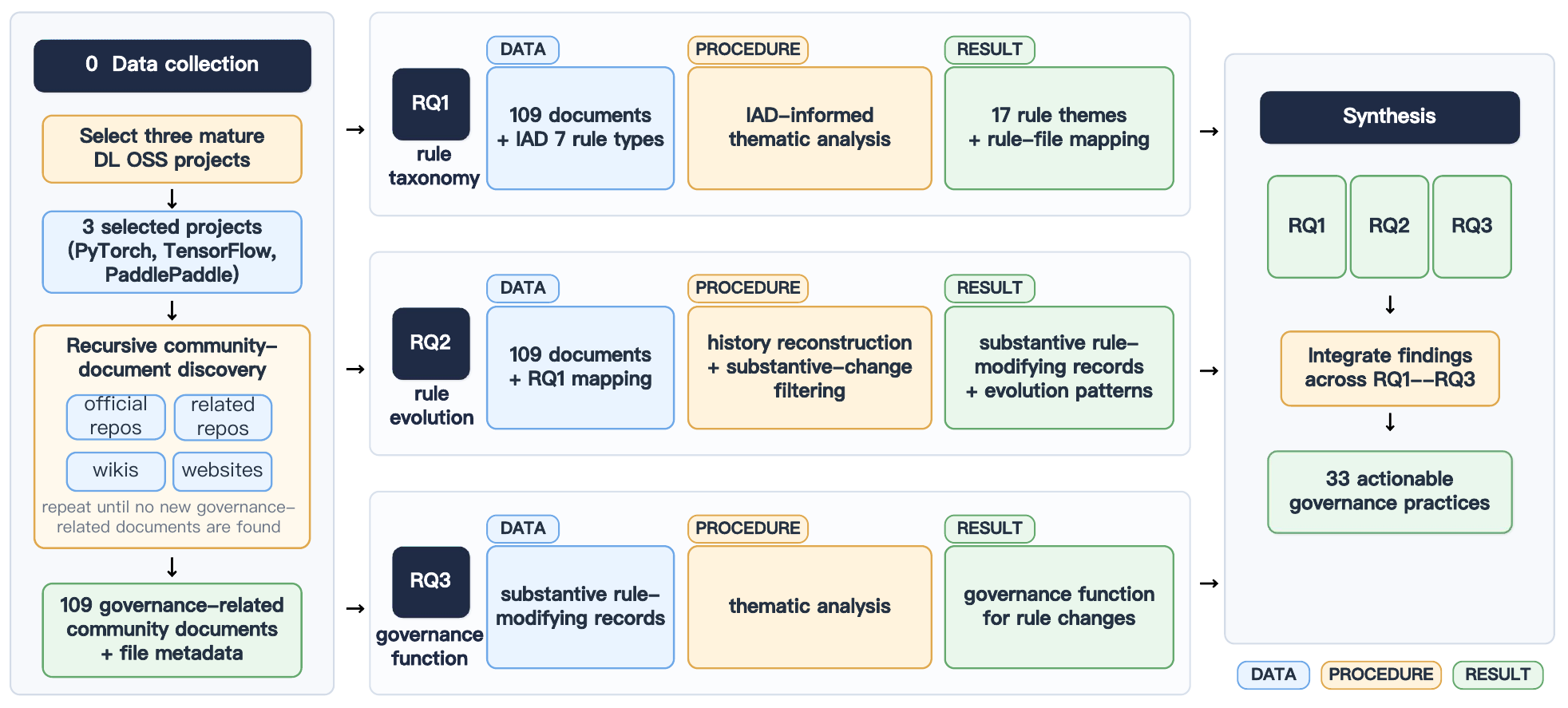}
  \caption{Overview of Study Design}
  \label{fig:methodology}
\end{figure}

\cref{fig:methodology} provides an overview of our methodology. We adopted a mixed-methods approach to address the three proposed RQs. The horizontal arrows from data collection to the three RQs indicate the use of the collected data, while those from the RQs to the synthesis stage indicate that their findings are integrated.

Section~\ref{ss:iadframework} introduces the IAD framework as the theoretical lens of this study, and Section~\ref{ss:data-collection} details the data collection process. Sections~\ref{s:rq1}, \ref{s:rq2}, and \ref{s:rq3} present the detailed methods and findings for the three RQs, respectively.

\subsection{Theoretical Lens: The IAD Framework}~\label{ss:iadframework}

To analyze OSS governance documents, we require a framework that can move beyond broad descriptions of governance domains and support the systematic classification of concrete rule provisions. Our objective is not only to identify whether projects address governance concerns such as roles, decision-making, participation, or coordination, but also to examine how these concerns are encoded in documentary rules and how such rules evolve over time. This requires an analytical lens that is sufficiently fine-grained to classify individual rule statements, and sufficiently general to be applied across heterogeneous community documents.

Existing OSS governance frameworks provide important foundations, but they are not primarily designed for this kind of rule-level documentary analysis. Markus et al.~\cite{markus2007governance} characterize governance through broad domains such as asset ownership, conflict resolution, and software development processes. While useful for identifying governance concerns, these domains are often cross-cutting when applied to individual documentary rules. For example, conflict resolution may be implemented through role definitions, decision procedures, information disclosure requirements, or sanctions. Similarly, O'Mahony et al.~\cite{o2007governance} identify normative principles such as independence, pluralism, representation, and decentralized decision-making. These principles help explain the values underlying governance arrangements but do not provide operational categories for classifying specific governance provisions. Chulani et al.~\cite{chulani2008software} distinguish between governance structures and execution mechanisms, yet this distinction remains too coarse-grained for organizing the diverse rules commonly found in OSS community documents.

Given this rule-level analytical objective, we adopt Ostrom’s Institutional Analysis and Development (IAD) framework as our theoretical lens~\cite{ostrom2009understanding}. Unlike governance frameworks that emphasize governance domains, organizational forms, or normative values, the IAD framework classifies institutional arrangements according to the functions they perform in structuring an action situation. Its rule categories are analytically distinguishable, can be operationalized at the level of individual documentary provisions, and can be applied consistently across different stages of governance evolution. These characteristics make IAD particularly suitable for our objective of systematically organizing OSS governance rules and tracing their evolution over time.

The suitability of IAD is further reinforced by the nature of OSS development itself. In OSS projects, decentralized contribution commonly coexists with project-level authority exercised by maintainers, repository owners, or governance bodies. Benkler even characterizes OSS as a “quintessential example” of commons-based peer production (CBPP)~\cite{cropf2008benkler}, in which individuals collectively produce a shared information resource through large-scale, decentralized, and peer-based collaboration. The IAD framework was originally designed to analyze how actors coordinate through shared rules, roles, and decision procedures in collective-action settings. Consequently, it provides a natural analytical lens for examining OSS governance.

In such settings, a central challenge is to manage collective-action dilemmas by aligning individual incentives and actions with collective goals across a distributed contributor base. OSS projects address this challenge through a combination of project-level authority, community norms, and documented rules. Although originally developed to explain the governance of common-pool resources, the IAD framework focuses more generally on how institutional arrangements structure interactions and address coordination problems. These roles, decision rights, and coordination arrangements are often encoded in community documents. Through the IAD lens, documentary governance rules can therefore be analyzed as mechanisms that shape participation, allocate authority, coordinate collaboration, and reduce transaction costs.

The analytical core of the IAD framework is the action situation, an interaction space in which participants occupying particular positions choose among available actions, exchange information, participate in collective decisions, and generate outcomes associated with costs and benefits. The seven IAD rule types specify different ways in which institutional arrangements structure this situation: position and boundary rules define roles and entry conditions; choice rules specify permitted, required, or prohibited actions; aggregation rules define how collective decisions are made; information rules regulate information availability and communication; scope rules define permissible outcomes; and pay-off rules assign rewards or sanctions.

In this study, we define the focal action situation as the OSS contribution workflow, including activities such as issue reporting, code submission, and pull-request review. We analyze documentary governance rules as institutional arrangements that structure this workflow. The seven IAD rule types are used as an analytical coding scheme for identifying textual provisions that instantiate each rule function in OSS community documents. This operationalization retains the original IAD categories while specifying OSS-specific observable evidence for each category. \cref{tab:iad} presents the original definitions of the seven rule types and their operationalization in the OSS context.

\begin{table}
  \caption{Operationalization of IAD Rule Types in OSS Governance}
  \label{tab:iad}
  \footnotesize
  \begin{tabular}{c p{4.5cm} p{5.5cm}}
    \toprule
    IAD Rule Type &
    \multicolumn{1}{c}{Original Description} &
    \multicolumn{1}{c}{Description for OSS Projects}\\
    \midrule
    Position Rules & Positions participants hold in an action situation & Roles defined within the OSS community and their associated authority, responsibilities, or support functions\\
    Boundary Rules & Rules that specify who is eligible to enter or hold a position  & Criteria and procedures that regulate how contributors enter, acquire, or leave community roles\\
    Choice Rules & Rules about what actions are permitted, required, or forbidden for each position & Documented prescriptions specifying what contributors are permitted, required, recommended, or prohibited to do during contribution and community participation activities \\
    Scope Rules & Rules that define permissible or expected outcomes & Standards defining the acceptable quality, style, or format of contribution outcomes\\
    Aggregation Rules & Rules about how collective decisions are made & Processes through which multi-participant decisions are reviewed, approved, or resolved, such as code integration or feature adoption\\
    Information Rules & Rules about information availability, flow, and access & Protocols determining where contributors communicate, what information they provide, and how project information is structured or accessed \\
    Pay-off Rules & Rules concerning rewards or sanctions tied to actions & Mechanisms for distributing recognition, rewards, or sanctions that shape contributor motivation and participation\\
    \bottomrule
  \end{tabular}
\end{table}

\subsection{Data Collection} ~\label{ss:data-collection}

To systematically analyze the governance rules in open source communities, we conducted an empirical study. Documents provide a concrete representation of abstract governance mechanisms, lowering the cognitive and communication barriers for contributors and reducing the coordination burden for maintainers. Therefore, we use community documents as the primary data source to identify and analyze the rules that structure community governance.

We selected three open-source deep learning frameworks as our empirical subjects: PyTorch, TensorFlow, and Paddle. This selection is aligned with our objective of examining governance rules that have been formalized and maintained in mature OSS communities. 

We focused on deep learning frameworks for two reasons. First, DL frameworks are widely adopted, rapidly evolving, and technologically complex, which makes coordination, contribution management, and rule formalization especially important. Second, restricting the cases to the same technical domain improves comparability across projects; otherwise, cross-domain differences could make it difficult to distinguish governance evolution from domain-specific development practices.

The selected projects satisfy three criteria. First, they are large-scale and actively maintained projects with substantial user and contributor communities, making them suitable for studying mature OSS governance. Second, they are hosted on GitHub and provide publicly accessible community documents, which enables systematic and reproducible document collection. Third, they provide a relatively complete set of governance-related materials. We operationalized this criterion by requiring GitHub-hosted repositories with the community-health artifacts highlighted by GitHub's Community Standards, such as \texttt{CONTRIBUTING.md}, \texttt{CODE\_OF\_CONDUCT.md}, and issue or pull-request templates, supplemented by official websites, wikis, and related repositories.

We adopted an iterative snowballing strategy to collect documents related to governance rules. Starting from each project's default repository (i.e., \texttt{pytorch/pytorch}, \texttt{tensorflow/tensorflow}, and \texttt{PaddlePaddle/Paddle}), we first screened seed documents such as \texttt{README.md}, \texttt{CONTRIBUTING.md}, and pull-request templates. In subsequent rounds, we followed URLs and cross-references in the documents identified from the previous round, and further screened the official document spaces discovered through these links, including GitHub Wikis, official websites, community repositories, RFC repositories, and official blogs. We repeated this process for three rounds and stopped when an additional pass produced no new governance-related documents satisfying our inclusion criteria.

During screening, we excluded documents that were marked as archived or deprecated, unrelated to project governance, purely technical, third-party or unofficial, duplicated or mirrored, non-document resources, or outside the project scope. The first author performed the initial screening according to these exclusion criteria, and documents with uncertain relevance were discussed by the first and second authors until agreement was reached. Notably, we retained official announcements and blog posts that define specific community activities or temporary governance mechanisms, such as PyTorch's Docathon activity. These sources were included because they often contain dynamic governance rules regarding participation and rewards that are not immediately codified in static documentation files.

We ultimately collected 109 governance-related documents. Our data collection covered the included sources up to December 31, 2024. Accordingly, the subsequent analyses characterize the documented governance of the three projects only up to this cutoff. \cref{tab:governance-doc-distribution} presents their distribution by project and source location. We provide this document index in the replication package.
Specifically, \textit{Project Repo} refers to each project's default code repository: \texttt{pytorch/pytorch}, \texttt{tensorflow/tensorflow}, and \texttt{PaddlePaddle/Paddle}. \textit{Official Website} refers to the project's official documentation or community website. \textit{Official Blog} refers to the project's official news or blog site. \textit{Wiki Pages} refer to the GitHub Wiki associated with the default repository. \textit{Community Repo} refers to a separate repository named \texttt{community} within the same GitHub organization, such as \texttt{tensorflow/community} and \texttt{PaddlePaddle/community}. \textit{RFC Repo} refers to a separate repository named \texttt{rfcs} within the same GitHub organization, such as \texttt{pytorch/rfcs}.
Except for official blog posts, the collected documents have accessible version histories through Git commits or Wiki revisions.

\begin{table}[htbp]
    \caption{Distribution of Governance-Related Documents by Project and Source}
    \label{tab:governance-doc-distribution}
    \footnotesize
    \centering
    \begin{tabular}{l|c c c c}
        \toprule
        \textbf{Source Location} & \textbf{PyTorch} & \textbf{TensorFlow} & \textbf{PaddlePaddle} & \textbf{Total} \\
        \midrule
        Project Repo & 12 & 11 & 12 & 35 \\
        Official Website & 4 & 9 & 12 & 25 \\
        Official Blog & 5 & 2 & 2 & 9 \\
        Wiki Pages & 15 & -- & -- & 15 \\
        Community Repo & -- & 14 & 9 & 23 \\
        RFC Repo & 2 & -- & -- & 2 \\
        \hline
        Total & 38 & 36 & 35 & 109 \\
        \bottomrule
    \end{tabular}
\end{table}

\section{RQ1: What specific governance rules are codified in OSS projects, how are they structured across documentation, and how do they vary across communities?}~\label{s:rq1}

This section reports how governance rules are identified, organized, and compared across the three OSS communities. We first describe our qualitative coding procedure, including how rule-bearing text segments were identified, how codes were grouped into rule themes, and how these themes were mapped to the IAD rule types. We then present the resulting taxonomy of rule themes and provide evidence examples from project documents. Next, we examine where these rule themes are codified across different document categories, using segment-level counts to show how rules are distributed across community artifacts. Finally, we compare the three projects to highlight both shared patterns and project-specific differences in documented governance rules.

\subsection{Method}

We followed a structured thematic coding process~\cite{cruzes2011recommended} to analyze the governance rules codified in the documents collected in \cref{ss:data-collection}.

First, we divided the documents into \textit{rule-bearing text segments}. A coded segment was defined as a coherent unit of text that described one specific governance arrangement or requirement, such as defining a role, assigning an issue, specifying a communication channel, or setting a contribution standard. For the official posts included in the dataset, we treated each complete post as one rule-bearing segment and assigned it one rule code, because each post described a specific community activity or temporary governance mechanism. Segment boundaries were determined by whether adjacent text addressed the same arrangement or requirement, rather than by paragraph boundaries. Therefore, a segment could range from a sentence, bullet point, or template field to multiple adjacent paragraphs. We included text specifying who may or should do what, under what conditions, through which channels, according to what standards, or with what consequences. We excluded contextual, duplicated, or purely informational text that did not itself specify a governance rule. Detailed exclusion criteria and examples are provided in the replication package.

Second, the first author conducted open coding by assigning each rule-bearing segment a \textit{rule code} that summarized the specific rule expressed by the segment. The second author reviewed the coded and excluded segments, preliminary code definitions, and code assignments. The two authors refined the code definitions and inclusion and exclusion criteria through multiple rounds of discussion. Persistent disagreements were referred to a third author.

Third, we iteratively grouped rule codes into higher-level \textit{rule themes} based on similarities in what the rules regulated. The level of abstraction was chosen to support comparison across projects and documents. For example, \textit{CLA Guidance}, \textit{Development Environment Setup}, and \textit{Submission Steps} were grouped under \textit{Contribution Submission Guidance}, as they all specify requirements for submitting contributions. Codes that already represented a coherent theme at this level, such as \textit{Code Review Process}, \textit{Code of Conduct}, and \textit{PR Size}, were retained directly as rule themes.

Fourth, we mapped each rule theme to one of the seven IAD rule types according to the definitions of these rule types: position, boundary, choice, aggregation, information, pay-off, and scope rules. For example, themes defining participant roles were mapped to position rules, those defining entry into or advancement between roles were mapped to boundary rules, and those defining acceptable contribution outcomes were mapped to scope rules. Potential overlaps were resolved through author discussion, and all identified themes could be assigned to one of the seven rule types.

The resulting coding hierarchy contains three levels: IAD rule types, rule themes, and rule codes. Rule themes represent recurring kinds of governance rules and serve as our primary unit of analysis, while rule codes describe more specific rules grouped under a broader theme. In total, we identified 17 rule themes and 36 lower-level rule codes. In the taxonomy and codebook, the leaf entries include both lower-level rule codes and rule themes that were not further divided.

Finally, we examined how rule themes were distributed across project documentation. Based on filename and primary purpose, one author developed a file-type codebook and classified the documents into nine categories: \textit{Overview Files}, \textit{Contributing Files}, \textit{Governance Structure Files}, \textit{Code of Conduct Files}, \textit{Security Files}, \textit{Issue Templates}, \textit{PR Templates}, \textit{RFC Templates}, and \textit{Posts}. A second author reviewed the codebook and category assignments, and disagreements were resolved through discussion. We then linked each coded segment to its document category and aggregated these links. The resulting Sankey diagram uses the number of coded rule-bearing segments as edge weights and retains all non-zero links between rule themes and document categories.

\subsection{Results}

\subsubsection{Taxonomy of Rule Themes Mapped to IAD Rule Types}

\begin{figure}[h]
  \centering
  \includegraphics[width=1\linewidth]{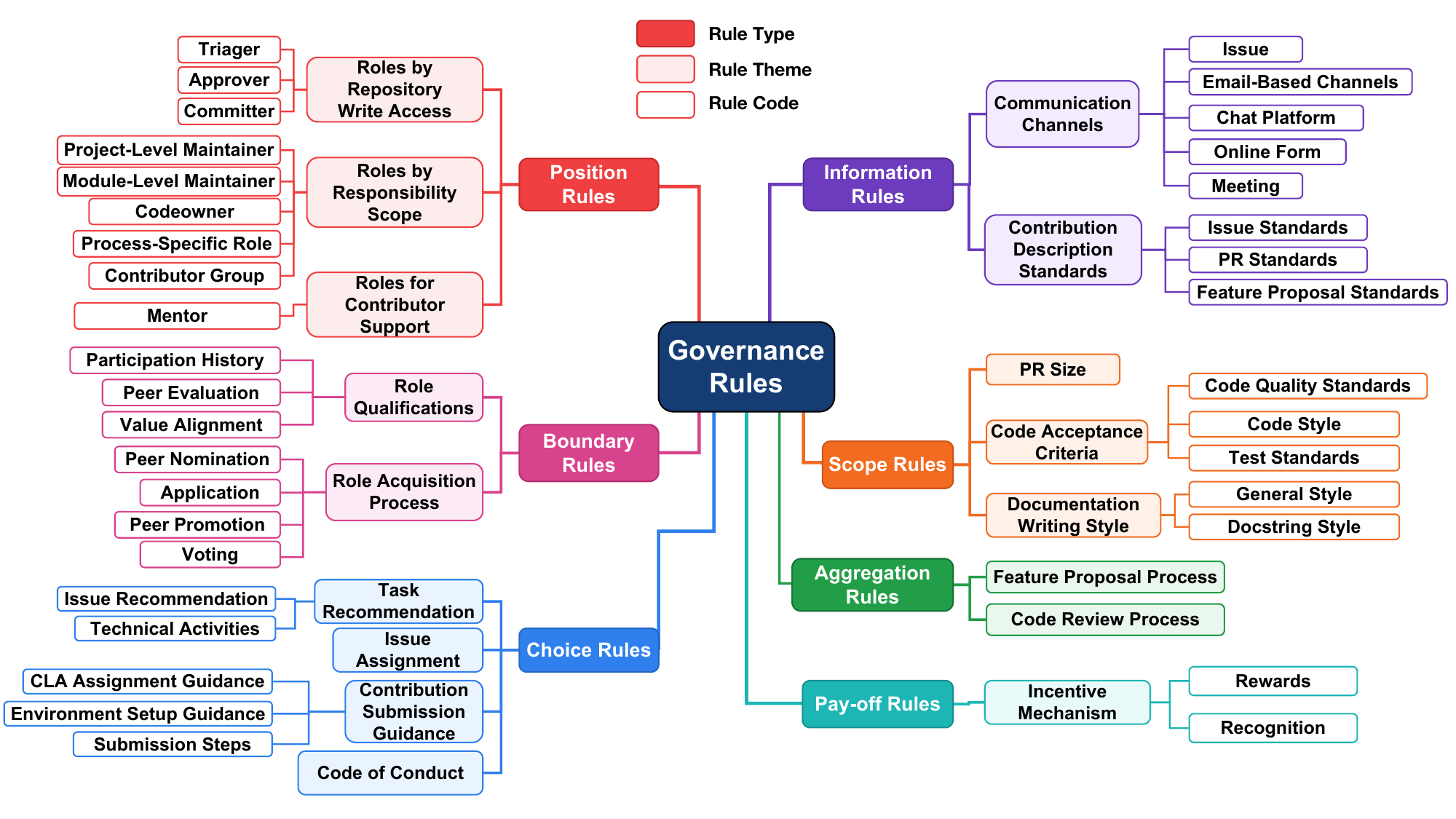}
  \caption{Taxonomy of Rule Themes Mapped to IAD Rule Types}
  \label{fig:taxonomy}
\end{figure}

\cref{fig:taxonomy} summarizes the taxonomy produced by the thematic coding process. The taxonomy organizes the 17 rule themes under the seven IAD rule types and shows lower-level rule codes where applicable. In the following subsections, we present the rule themes by IAD rule type and describe how they regulate participation and coordination in OSS communities. We include representative examples in the text to illustrate how themes were interpreted; the complete mapping from source excerpts to rule codes, rule themes, and IAD rule types is provided in the replication package.

\textbf{Position Rules} define the roles participants can hold and the authority, responsibility, or support functions associated with those roles. We identified three rule themes: \textbf{Roles by Repository Write Access}, \textbf{Roles by Responsibility Scope}, and \textbf{Roles for Contributor Support}. These themes are not mutually exclusive categories of participants; instead, they capture different dimensions used in project documents to define roles. Roles by Repository Write Access emphasize technical permissions over repository operations, such as triaging, approving, or merging changes. Roles by Responsibility Scope emphasize the domain, artifact, or governance process for which participants are responsible, such as a project, module, file path, SIG, or RFC review process. For example, TensorFlow defines approvers through repository permissions while also defining CODEOWNERS, SIG members, and RFC reviewers through responsibility scopes; the same participant may be described by both dimensions. Roles for Contributor Support focus on supporting contributor engagement. Paddle introduces \textit{mentors} to provide guidance and facilitate the smooth onboarding of newcomers.

\textbf{Boundary Rules} regulate how participants become eligible for, acquire, or leave community roles. We identified two rule themes: \textbf{Role Qualifications} and \textbf{Role Acquisition Process}. Role Qualifications define eligibility conditions, such as sustained contribution history, demonstrated contribution quality, peer evaluation, or alignment with project values. For example, PyTorch describes maintainer advancement as merit-based, while Paddle requires committer candidates to have completed high-quality pull requests and to be recommended by existing committers. Role Acquisition Process defines the procedures through which roles are obtained, including peer nomination, formal application, voting, and promotion by existing role holders. Overall, boundary rules tend to be more explicit for roles with repository authority or broad governance responsibility, such as committers and maintainers, than for support-oriented roles such as mentors. Rather than indicating the absence of governance, this variation suggests that projects codify role boundaries more explicitly when the role carries greater authority over code integration, review decisions, or community governance.

\textbf{Choice Rules} specify what contributors are permitted, required, recommended, or prohibited to do during contribution and community participation. We identified four rule themes: \textbf{Task Recommendation}, \textbf{Issue Assignment}, \textbf{Contribution Submission Guidance}, and \textbf{Code of Conduct}. Task Recommendation helps contributors identify suitable work, for example through labels such as ``good first issue'' or organized contribution activities. Issue Assignment coordinates commitment to a specific task, such as asking contributors to comment on an issue before starting work to avoid duplicated effort. Contribution Submission Guidance specifies procedural prerequisites for valid contributions, including CLA guidance, development environment setup, and pull request submission steps. The Code of Conduct theme captures documented behavioral obligations and prohibitions for community interaction. These themes may also operate sequentially rather than independently: \textit{Task Recommendation} helps contributors discover suitable work, while \textit{Issue Assignment} coordinates commitment after a task has been selected. Together, the four themes reduce coordination costs by making expected contributor actions explicit.

\textbf{Aggregation Rules} define how multi-participant decisions are reviewed, approved, or resolved, as these decisions may influence the software quality or even affect the overall direction of the project. We identified two rule themes: \textbf{Code Review Process} and \textbf{Feature Proposal Process}. Code Review Process governs the daily integration of code, aiming to streamline asynchronous interactions between authors and reviewers. For example, to mitigate delays caused by reviewer workloads, PyTorch establishes clear expectations, encouraging reviewers to respond within 24 hours and empowering authors to proactively follow up on stagnant PRs. Feature Proposal Process governs higher-impact decisions, such as introducing new features, changing APIs, or making architectural changes. Across the studied projects, this process is often codified through Requests for Comments (RFCs) or similar design-proposal procedures. These procedures require contributors to justify the motivation, design choices, maintainability, and potential impact of a proposal before implementation, thereby aligning stakeholders early and reducing the risk of costly rework.

\textbf{Information Rules} define where contributors communicate, what information they should provide, and how project information is structured or accessed. These rules help developers maintain coordinated progress and prevent misunderstandings. We identified two rule themes: \textbf{Communication Channels} and \textbf{Contribution Description Standards}. 
Communication Channels specify the venues contributors should use for different interactions. For instance, PyTorch directs user questions to \textit{forums}, bug reports to \textit{GitHub issues}, and real-time build support to \textit{meetings}; meanwhile, TensorFlow utilizes \textit{mailing lists} for announcements, and Paddle developers leverage \textit{online chat rooms} for instant collaboration. We coded a resource as a communication channel only when it prescribed where contributors should submit information, ask questions, report problems, or provide feedback. Contribution Description Standards specify what information contributors should provide in issues, pull requests, or RFCs. For example, all three projects adopt description standards for issues, pull requests, and feature proposals to guide contributors in presenting their input clearly.

\textbf{Scope Rules} define the acceptable outcomes of contributor actions, specifying what constitutes a "good" contribution to minimize misalignment and future maintenance costs. We identified three rule themes: \textbf{PR Size}, \textbf{Code Acceptance Criteria}, and \textbf{Documentation Writing Style}. PR Size rules regulate the granularity of submissions, for example by discouraging overly large pull requests or trivial edits. Code Acceptance Criteria define the technical quality requirements that contributions must satisfy. This involves adhering to \textit{code quality standards} (e.g., maintaining API compatibility of PyTorch), following strict \textit{code styles} (all three projects adopt Google's Python and C++ style guides), and meeting \textit{test standards} (e.g., Paddle mandates relevant tests with sufficient coverage for functional changes). Documentation Writing Style defines standards for non-code artifacts. This includes \textit{general style} guidelines for Markdown and other documentation files, as well as \textit{docstring style} rules, such as TensorFlow's standards for API documentation and docstrings.

\textbf{Pay-off Rules} regulate the distribution of recognition, rewards, or sanctions associated with contributor actions. We identified one governance rule theme: \textbf{Incentive Mechanism}. In the studied documents, this theme appears mainly in two forms. The first is \textit{recognition}, where projects provide non-material acknowledgment, such as certificates, public acknowledgment, contributor spotlights, awards, or other honor-based recognition. The second is \textit{rewards}, where projects provide tangible incentives, such as monetary prizes, gifts, or internship opportunities. For example, the PyTorch and TensorFlow documents we analyzed mainly codify non-material recognition, such as public acknowledgment, contributor spotlights, or awards, whereas the Paddle documents also codify tangible incentives, such as monetary prizes and internship opportunities. These incentive mechanisms make desirable forms of participation more visible and help projects signal which contributions they seek to encourage.

Our taxonomy both refines and extends prior semantic classifications of OSS governance. Prior studies identified broad topics such as mentorship, voting, mailing lists, documentation, testing, forms, and issues~\cite{yin2022open, chakraborti2024we}. By mapping documented rules to the IAD rule types, our taxonomy clarifies how these topics regulate collaboration, such as defining roles, controlling role entry, structuring decisions, regulating information exchange, specifying acceptable outcomes, or configuring incentives.

It also identifies themes that were not represented as standalone categories in either prior classification. In particular, \textit{PR Size} captures contribution granularity as a constraint on acceptable outcomes, while \textit{Incentive Mechanism} captures rewards and recognition as pay-off rules. Other themes make finer distinctions within broader prior topics, such as separating role qualifications from role-acquisition procedures and distinguishing documentation-writing standards from general documentation activity. \cref{tab:prior-work-examples} summarizes representative additions and refinements.

\begin{table}[t]
\centering
\caption{Rule themes newly identified or more explicitly distinguished relative to prior semantic classifications.}
\label{tab:prior-work-examples}
\small
\begin{tabular}{p{0.22\linewidth} p{0.24\linewidth} p{0.44\linewidth}}
\toprule
\textbf{Rule theme} &
\textbf{Closest prior coverage} &
\textbf{Added distinction} \\
\midrule

Role Qualifications and Role Acquisition Process &
Committers/Commits; General Voting &
Separates eligibility conditions, such as participation history and peer evaluation, from procedures for acquiring a role, such as nomination, application, promotion, and voting. \\

Contribution Description Standards &
Forms; Issues; Proposals/Resolutions &
Unifies issue, pull-request, and proposal templates as information rules specifying what information contributors must provide. \\

Documentation Writing Style &
Documentation &
Distinguishes standards for documentation content and docstrings from general documentation-related activity. \\

PR Size &
No direct counterpart &
Identifies contribution granularity as an explicit limit on acceptable contribution outcomes. \\

Incentive Mechanism &
No direct counterpart &
Identifies rewards and recognition as pay-off rules that configure the material or reputational benefits associated with participation. \\

\bottomrule
\end{tabular}
\end{table}

\subsubsection{Distribution of Rules across Community Documents}

\begin{figure}[h]
  \centering
  \includegraphics[width=1.0\linewidth]{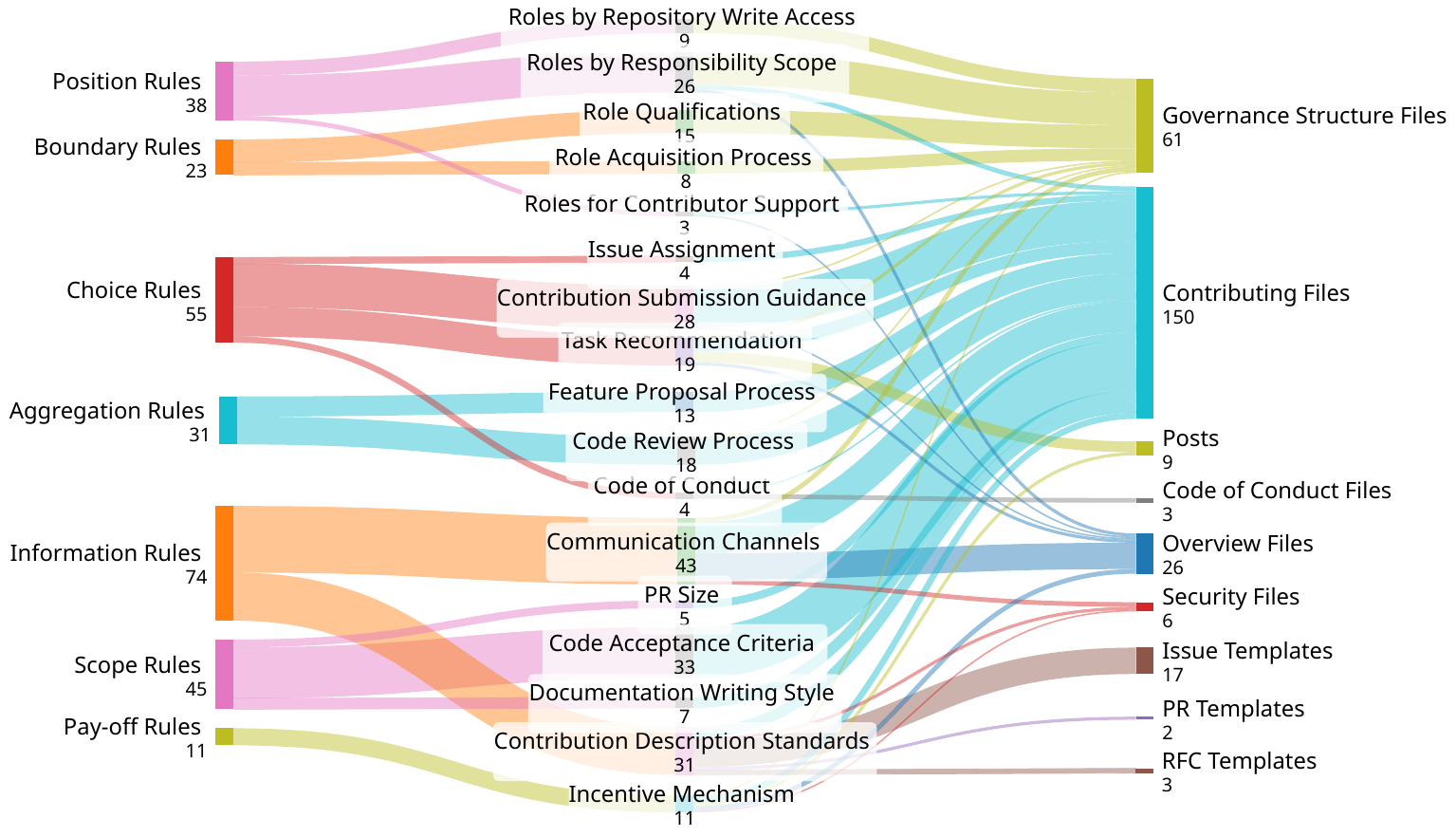}
  \caption{Distribution of Governance Rule Themes Across Community Document Categories}
  \label{fig:sankey}
\end{figure}

\cref{fig:sankey} shows how the identified rule themes are distributed across different categories of community documents. In this figure, the width of each flow represents the number of coded rule-bearing text segments. Because the same governance rule theme can be codified in multiple types of documents, we preserve all non-zero links between rule themes and document categories.

The distribution reveals that different document categories play distinct roles in making governance rules visible to contributors. \textit{Governance Structure Files} primarily contain rules about organizational structure and role boundaries. For example, most segments related to \textit{Roles by Repository Write Access}, \textit{Roles by Responsibility Scope}, \textit{Role Qualifications}, and \textit{Role Acquisition Process} appear in this document category. This pattern suggests that rules defining who can hold specific roles, how authority is allocated, and how contributors move into formal positions are usually centralized in documents that describe project governance structures.

By contrast, \textit{Contributing Files} serve as the main location for operational rules that guide day-to-day participation. These files contain many segments related to \textit{Contribution Submission Guidance}, \textit{Code Review Process}, \textit{Feature Proposal Process}, \textit{Code Acceptance Criteria}, \textit{Documentation Writing Style}, and \textit{PR Size}. This placement is consistent with the function of contributing documents as workflow manuals: they tell contributors how to prepare, submit, revise, and validate their contributions.

Information-related rules are more distributed. \textit{Communication Channels} appears across \textit{Overview Files}, \textit{Contributing Files}, \textit{Governance Structure Files}, and \textit{Security Files}, reflecting the fact that communication rules are embedded in both general orientation documents and task-specific procedures. For instance, overview documents often introduce forums, mailing lists, or other community entry points, while security files specify dedicated reporting channels for vulnerabilities. Similarly, \textit{Contribution Description Standards} are distributed across \textit{Contributing Files}, \textit{Issue Templates}, \textit{PR Templates}, and \textit{RFC Templates}. These templates operationalize information rules by prompting contributors to provide required contextual information when submitting issues, pull requests, or design proposals.

Some rule themes are intentionally documented in more dynamic or community-facing formats. \textit{Task Recommendation} appears not only in contributing documents but also in posts, where projects advertise contribution opportunities, campaigns, or newcomer-friendly activities. \textit{Incentive Mechanism} is also distributed across posts, overview documents, contributing files, and governance structure files, indicating that reward- and recognition-related rules are often communicated through announcements or community-facing materials rather than only through static governance documents.

Overall, the distribution shows that governance rules are not stored in a single location. Rule themes related to roles and role transitions are mostly documented in Governance Structure Files, while themes related to contribution submission, review, and acceptance criteria are concentrated in Contributing Files and templates. In contrast, information- and incentive-related themes are more widely distributed across Overview Files, Contributing Files, Security Files, Governance Structure Files, and Posts. This distribution also highlights a navigation challenge for newcomers. The collected documents suggest multiple entry points into the rule ecosystem, so understanding the full set of documented governance rules may require contributors to move across multiple linked document categories rather than reading a single governance file.

\subsubsection{Comparison across Projects} ~\label{sec:diff-in-rules}
The three projects show stronger commonality in rule themes that guide day-to-day contribution work. We use \textit{operational rule themes} to refer to themes that directly guide contributors in routine contribution activities, such as contribution submission guidance, code review processes, feature proposal processes, communication channels, contribution description standards, and code acceptance criteria. These themes appear across all three projects, and many of their lower-level rule codes are also shared. For example, all three projects document CLA-related guidance, development environment setup, submission steps, code review processes, feature proposal processes, issue or pull-request templates, and code acceptance criteria. Their concrete procedures, communication media, and template details still vary, but the underlying contribution functions are commonly codified. This suggests that, despite differences in organizational context, the projects share a common need to make routine contribution activities legible and actionable for contributors.

By contrast, the more visible project-specific differences appear in themes related to roles, role-entry conditions, and incentives. These differences concern not only implementation details, but also which role forms, entry pathways, and incentive forms are codified in public documents. For position rules, Paddle documents several forms of roles, including repository-authority roles such as committers, triagers, and approvers; contributor-support roles such as mentors; and community- or process-oriented structures such as SIGs, PFCC-related roles, and process-specific roles. TensorFlow documents repository or review-related roles such as maintainers or approvers, as well as codeowners, contributor groups, SIG-related structures, and RFC-related process roles. PyTorch documents a more concise set of role-related rules in our corpus, with greater emphasis on project or module maintainers and codeowners. These differences suggest that projects vary in which aspects of responsibility, authority, and coordination they make explicit in public documentation.

Boundary rules also vary across projects in the pathways they codify for entering or changing roles. Paddle documents application-, contribution-, recommendation-, and voting-related procedures for some roles. PyTorch documents participation-history requirements, peer nomination or voting procedures, and value-alignment criteria for maintainer-related roles. TensorFlow's boundary rules are less explicit for approver-related advancement, but more detailed in the context of SIG participation and formation. These patterns show that role-entry conditions are not codified through a single common pathway across projects; instead, public documents combine different criteria such as contribution history, peer evaluation, application, promotion, voting, and value alignment.

Pay-off rules are codified in more uneven forms across the three projects. All three projects contain recognition-oriented rules or materials, such as public acknowledgment of contributors. Paddle additionally codifies reward-oriented rules, including task- or program-based incentives. In contrast, the PyTorch and TensorFlow materials in our corpus primarily provide recognition-oriented examples. This difference indicates that contributor incentives are made explicit in different forms across the three projects' public artifacts.

Overall, the comparison suggests that the three projects commonly codify rule themes for routine contribution work at the theme and rule-code levels, while differing more visibly in how they codify role structures, role-entry conditions, and contributor incentives. These differences are best understood as variation in what public documents make explicit to contributors: some rules guide everyday contribution activities, whereas others define who can hold particular roles, how participants move into those roles, and what forms of recognition or reward are attached to contribution.

\begin{findingbox}{Summary for RQ1}
(1) \textbf{Content and structure}: We identified 17 rule themes mapped to the seven IAD rule types, covering roles, role-entry conditions, contributor actions, decision-making, information exchange, acceptable contribution outcomes, and incentives.
(2) \textbf{Document distribution}: Rule themes are distributed across multiple community artifacts. Role-related themes are mostly codified in governance structure files, contribution workflow and quality-related themes in contributing files and templates, and information or incentive-related themes across a wider range of overview, procedural, security, and announcement materials.
(3) \textbf{Cross-project variation}: The projects share many rule themes for routine contribution work such as submission guidance and code review, but differ more visibly in how they document role structures, role-entry pathways, and contributor incentives.
\end{findingbox}

\section{RQ2: How do governance rules evolve throughout the project lifecycle in terms of timing, frequency, and documentation structure?}~\label{s:rq2}

This section examines how governance rule themes evolve over time across the three OSS communities. We first define three longitudinal metrics used in the analysis: first appearance in documentation, modification frequency, and documentation splitting. We then describe how we reconstructed traceable documentation histories, identified substantive rule-modifying records, and recorded which logical files contained substantive rule-bearing text associated with each rule theme. Finally, we present the results along these three dimensions to show when rule themes first appeared in the collected documentation, how frequently their documented mechanisms were revised, and how widely their rule-bearing text spread across distinct logical files over time.

\subsection{Method}

To investigate the longitudinal evolution of governance rule themes, we analyzed the version histories of the governance-related documents identified in \cref{ss:data-collection}. Following the taxonomy developed in RQ1, we used rule themes as the main unit of longitudinal analysis and examined when they first appeared in the collected documentation, how frequently their documented mechanisms changed, and how their rule-bearing text spread across documentation files.

\subsubsection{Evolutionary Metrics}

We derived three metrics to characterize the longitudinal evolution of each rule theme:

\begin{itemize}
    \item \textbf{First Appearance in Documentation}: the timestamp at which substantive rule-bearing text associated with a rule theme first appeared in a project's collected documentation.

    \item \textbf{Modification Frequency}: the cumulative number of substantive project-level rule-modifying records associated with a rule theme, reflecting how frequently its documented mechanism changed.

    \item \textbf{Documentation Splitting}: the cumulative number of distinct logical files in which a rule theme had appeared by each year, reflecting its historical spread across documentation artifacts rather than the number of files active in a given year.
\end{itemize}

\subsubsection{Commit History Extraction}

The version history of OSS documentation is often discontinuous because files may be renamed, moved across directories, or migrated to separate repositories. To reduce the risk of overlooking earlier rule-bearing content, we employed a recursive retrieval and stitching strategy to reconstruct the traceable commit history of each governance-related document.

We first retrieved the commit logs of the current governance-related documents following established research methods~\cite{ihara2014early, sinha2011entering}. We then inspected the earliest available commit date of each file. If a file's history began substantially later than the project's creation date (e.g., a file appearing in 2021 for a project created in 2016) or coincided with a known repository migration event, we flagged it as a potential history discontinuity.

For files with potential history discontinuities, we traced predecessor paths in older directories or repositories using two strategies. First, we followed explicit migration clues, such as redirection URLs, repository migration notes, and migration-related pull requests. For example, some deprecated PyTorch documentation pages pointed to their new Wiki locations. Second, when explicit clues were unavailable or led to incomplete histories, we searched relevant project and documentation repositories using historical directory structures, file names, and distinctive text fragments from the documents. For manually identified predecessor candidates, we checked whether the candidate and current paths showed continuity in file names, document purpose, rule-bearing content, and commit timing. For example, in Paddle, the initial Git extraction traced some contribution-related documents only back to 2022, while repository evidence showed that related documentation had been migrated from the main Paddle repository to the Paddle docs repository around 2018. We therefore used intermediate path changes, filenames, and overlapping contribution-guidance content to trace predecessor paths across the two repositories.

When a predecessor-current link was established, we retrieved the commit logs of the predecessor path and appended them to the history of the current file. If no predecessor path could be identified, we used the earliest available commit of the current file as the start of its traceable history. We repeated this process recursively until no further predecessor path could be identified.

Using this approach, we processed the identified history discontinuities and additionally traced predecessor paths or directories beyond the current document paths. This reconstruction produced a dataset of 1,754 commits for subsequent filtering. The dataset represents the traceable version history of the governance-related documents included in our analysis.

\subsubsection{Construction of Rule-Modifying Records}

Our analysis used two types of time-stamped evidence. For continuously maintained documents, we used commit histories to identify when rule-bearing text was added, removed, or revised. For official posts included in RQ1, we used their publication dates because they function as dated governance records. We use the term \textit{record} to denote a time-stamped instance in which substantive rule-bearing text associated with one rule theme appeared or changed. 

For continuously maintained documents, records were constructed from commit diffs. The 1,754 retrieved commits included many documentation updates that did not change governance rules. We therefore manually inspected each commit diff and identified substantive rule-modifying records in each governance-related file. We retained a record only when it substantively changed the documented rule mechanism. Edits that preserved the underlying mechanism, such as wording or formatting changes, document relocation or reorganization, updates to concrete instances, and purely informational or technical changes, were excluded. When a commit contained both substantive and non-substantive edits, only the substantive change was coded. If one commit affected multiple rule themes or files, separate file--rule-theme records were created. For posts, when a post introduced substantive rule-bearing text associated with a rule theme, we treated its publication date as the record date for the corresponding project and rule theme. This process produced 335 file- or post-level rule-theme records.

We then used these records differently across the three metrics. For the first-appearance and modification-frequency metrics, we further grouped these records by project and rule theme and assessed whether each represented a substantive project-level change. This step avoided double-counting cases in which rule-bearing text was copied, moved, or reworded across files without changing the underlying rule. 
The first author conducted the file-level filtering and project-level aggregation; the criteria and borderline cases were discussed with the second author and refined accordingly. This process resulted in 265 project-level substantive rule-modifying records.

For documentation splitting, we retained the file-level perspective and counted the distinct logical files in which substantive rule-bearing text for each theme had appeared over time. This allowed us to capture the spread of governance documentation across artifacts. Clear renames and migrations were normalized as the same logical file, whereas topic-specific splits were retained as distinct files when they formed specialized documentation artifacts.

\subsection{Results}

\subsubsection{Timing of First Appearance in Documentation}
\begin{figure}[h]
  \centering
  \includegraphics[width=1\linewidth]{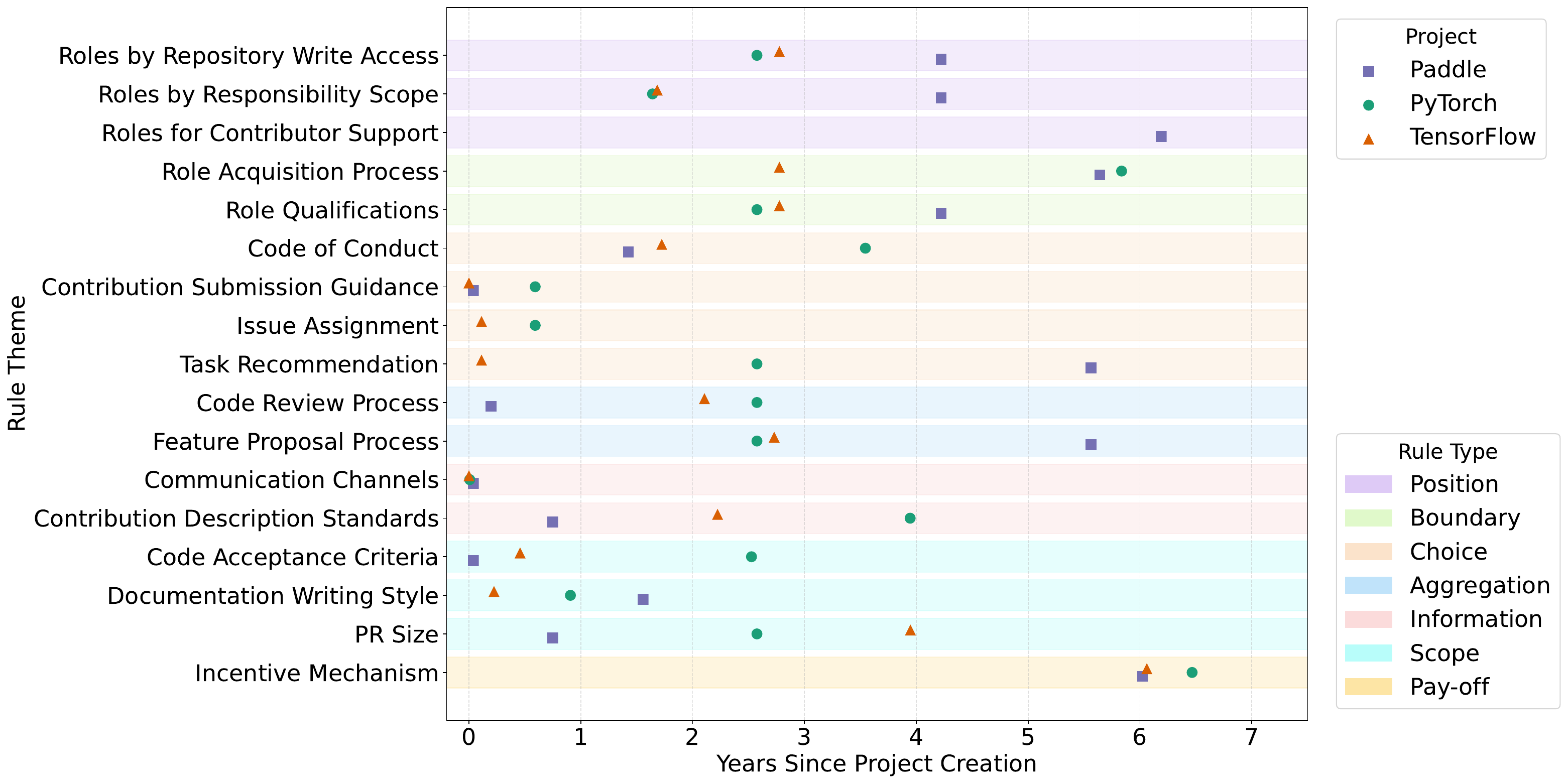}
  \caption{First Appearance of Each Rule Theme in Project Documentation Relative to Project Creation}
  \label{fig:rule_first_occurrence}
\end{figure}

\cref{fig:rule_first_occurrence} illustrates when each rule theme first appeared in the documentation of the three projects, relative to each project's creation date. In \cref{fig:rule_first_occurrence,fig:rule_change_occurrence}, vertical offsets are used only to separate overlapping project markers and do not encode additional values. Overall, the studied projects show an ``\textbf{operation first, structure later}'' tendency in their documentation: operational rule themes generally appeared earlier in the collected documentation, whereas role- and boundary-related themes often appeared after basic contribution workflows had already been documented.

During the initial phase, especially within the first two years, rule themes mapped to choice, aggregation, information, and scope rule types were among the first to appear in the projects' documentation. These themes concern practical coordination needs in daily development, such as communication channels, contribution submission, feature proposals, code review, and code acceptance criteria.

By contrast, many rule themes mapped to position and boundary rule types appeared later in the studied projects' documentation. These themes define community roles, responsibility scopes, role qualifications, and role acquisition processes, and their later documented occurrence indicates that formal role structures were publicly codified after basic contribution workflows had already been documented.
The rule theme mapped to pay-off rules, \textit{Incentive Mechanism}, appeared later in the collected documentation than most operational themes. This lag indicates that formal incentives and recognition were publicly codified only after the studied projects had already documented basic contribution and review workflows.

Across projects, the timing of first appearance in documentation varied. A broader range of rule themes appeared in TensorFlow's documentation within its first five years, while PyTorch followed a similar but more gradual trajectory. Paddle showed a more dispersed pattern, with several structural rule themes appearing later in its documented history. These differences are consistent with variation in project scale, contributor base, and organizational context, which may shape when communities formalize different kinds of governance documentation.

\subsubsection{Frequency and Timing of Modifications}
\begin{figure}[h]
  \centering
  \includegraphics[width=1\linewidth]{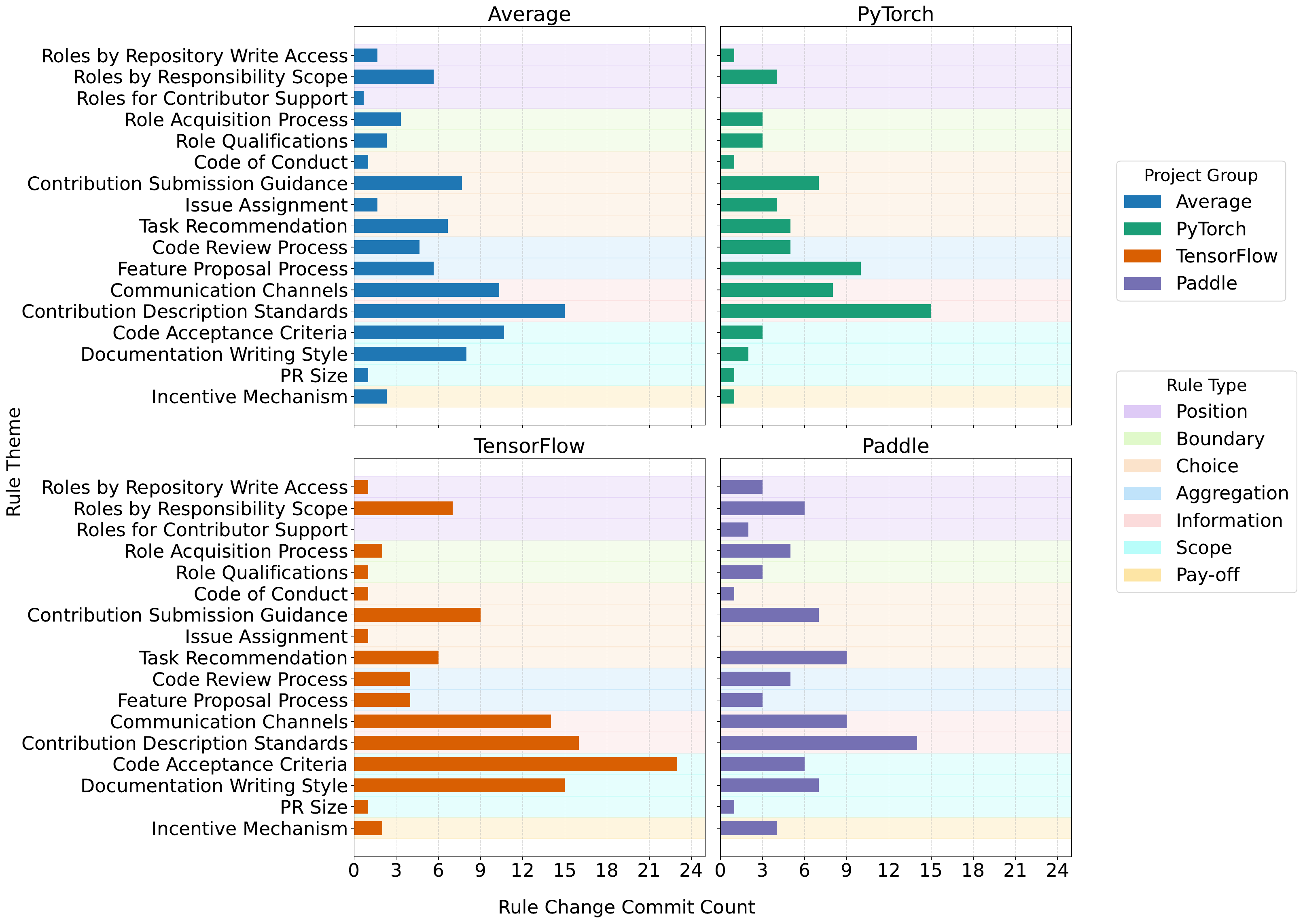}
  \caption{Modification Frequency of Rule Themes Across the Three Projects and Their Average}
  \label{fig:rule_change_frequency_comparison}
\end{figure}

\cref{fig:rule_change_frequency_comparison} shows how often each rule theme was modified. The results highlight a clear evolution trend: ``\textbf{stable governance structure, continuously refined operational practices}''. The most frequently modified rule themes were related to contribution content and quality. Specifically, \textit{Contribution Description Standards} (e.g., templates for PRs and issues) and \textit{Code Acceptance Criteria} (e.g., coding style) had the highest number of changes. Process-oriented rule themes, such as \textit{Contribution Submission Guidance} and \textit{Code Review Process}, were also revised frequently. This frequent updating suggests that communities repeatedly refined their daily workflows to accommodate changing technical and coordination needs. In contrast, rule themes mapped to position, boundary, and pay-off rule types were modified less frequently after their initial codification.

\begin{figure}[h]
  \centering
  \includegraphics[width=1\linewidth]{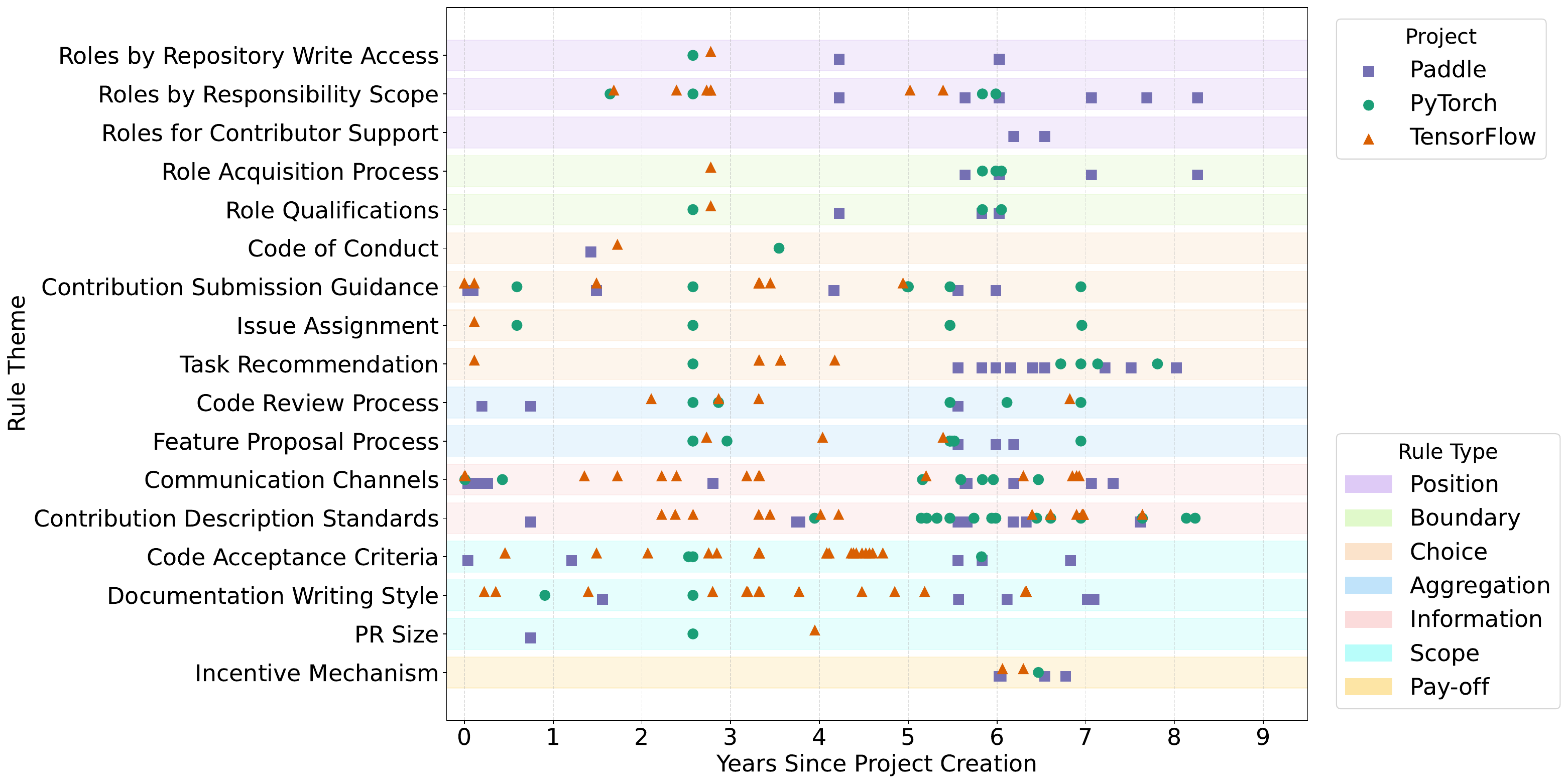}
  \caption{Rule Theme Modification Occurrences Relative to Project Creation}
  \label{fig:rule_change_occurrence}
\end{figure}

The timing of these changes (\cref{fig:rule_change_occurrence}) further reveals two different patterns. 
Role- and boundary-related themes were modified less frequently overall, and their changes were often clustered around later stages or major project events. For instance, PyTorch rarely updated its role-related rule themes, and the observed updates were concentrated around later milestones such as the release of version 1.0 in 2018 and joining the Linux Foundation in 2022.
By contrast, operational rule themes mapped to choice, information, aggregation, and scope rule types were revised more gradually over time. For example, TensorFlow expanded its contribution requirements from basic naming conventions to more detailed guidance on API compatibility and testing.

At the project level, the most frequently modified rule themes within each project reflected different operational adjustment priorities.
In PyTorch, \textit{Contribution Description Standards} accounted for a large share of modifications, including updates to issue and PR templates that clarified what information contributors should provide. In TensorFlow, the most frequently modified themes were concentrated around technical quality and documentation quality, especially \textit{Code Acceptance Criteria}, \textit{Documentation Writing Style}, and \textit{Contribution Description Standards}. These changes repeatedly refined requirements related to coding style, testing, API compatibility, and documentation conventions. Paddle showed more frequent modifications to \textit{Contribution Description Standards}, \textit{Contribution Submission Guidance}, \textit{Communication Channels}, and \textit{Documentation Writing Style}, reflecting repeated adjustments to submission steps, required contributor information, communication venues, and documentation-quality expectations. These patterns suggest that although structural rule themes changed infrequently across the projects, operational rule themes were adjusted in project-specific ways.

\subsubsection{Documentation Splitting and Specialization}
\begin{figure}[h]
  \centering
  \includegraphics[width=1\linewidth]{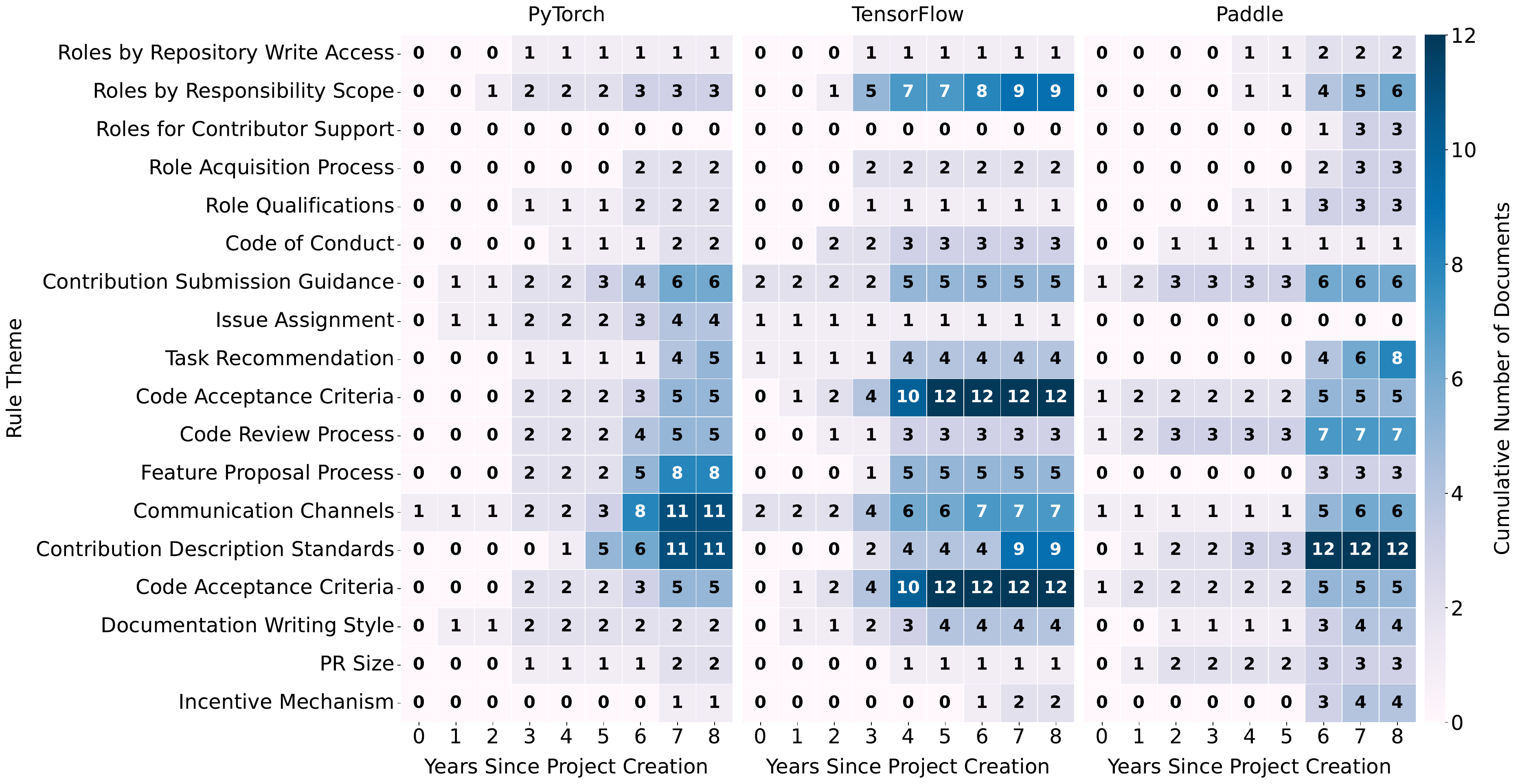}
  \caption{Cumulative Number of Distinct Logical Files Containing Each Rule Theme Over Time}
  \label{fig:rule_year_heatmap_all_projects}
\end{figure}

\cref{fig:rule_year_heatmap_all_projects} visualizes documentation splitting for each rule theme over time. Each cell reports the cumulative number of distinct logical files in which substantive rule-bearing text associated with a given rule theme had appeared by a given year since project creation. This metric captures the historical spread of rule themes across documentation artifacts rather than the exact number of active files in each year.
In the studied projects, many rule themes were initially concentrated in a small number of general documents, such as \texttt{CONTRIBUTING.md}, and later appeared across multiple task-specific or role-specific files. We interpret this increase in file distribution as documentation splitting, which often reflects the specialization of governance documentation rather than necessarily indicating disorder. The increase was not uniform across projects or rule themes: some operational rule themes continued to spread across additional files in later years, whereas several structural, conduct-related, and incentive-related themes remained concentrated in a small number of documents.

Across the three projects, \textit{Contribution Description Standards} showed the strongest documentation splitting, appearing in at least nine distinct logical files in each project by the end of the observation window. As shown in the document-category mapping in RQ1, this rule theme was frequently codified in contributing files. Because different contribution workflows require different information from contributors, the theme accumulated across multiple templates and contribution guides. In contrast, \textit{Code of Conduct} and \textit{Incentive Mechanism} showed limited documentation splitting. They were usually codified in a small number of dedicated documents, such as code-of-conduct files, governance files, or community program pages, rather than being repeatedly distributed across workflow-specific templates and contribution guides.

The specific forms of documentation splitting differed across projects. PyTorch adopted a workflow-oriented structure, separating guidance for reporting bugs, submitting PRs, asking questions, and proposing features into distinct documents. These workflow-specific files often combine the relevant communication channels, required information, and submission steps for a particular contributor task. TensorFlow developed more specialized documentation around SIG management, feature proposals, design review, and technical acceptance criteria. This pattern suggests an emphasis on coordinating technical decision-making and review processes. Paddle showed a later increase in specialized documentation, especially around contribution submission, code review, documentation quality, docstring style, and error-message writing. Overall, these differences suggest that documentation splitting was shaped by each project's operational priorities rather than following a single uniform structure.

\begin{findingbox}{Summary for RQ2}
(1) \textbf{Timing} (Operation First, Structure Later): In the studied projects, operational rule themes appeared earlier in the collected documentation, whereas many structural rule themes appeared later, after basic contribution workflows had already been documented.
(2) \textbf{Frequency} (Stable Structural Themes, Continuously Refined Operational Themes): Operational rule themes were modified more frequently to accommodate workflow and quality-control adjustments, whereas structural rule themes changed less often after their initial codification.
(3) \textbf{Structure} (Documentation Splitting and Specialization): Governance rule themes accumulated across multiple distinct logical files over project history. This accumulated file-level distribution suggests that governance documentation became more specialized as coordination needs became more differentiated.
\end{findingbox}

\section{RQ3: What governance functions are reflected in the formulation and modification of governance rules?} \label{s:rq3}

This section examines the governance functions reflected in the 265 project-level substantive rule-modifying records identified in RQ2. We first describe the coding procedure and the treatment of ambiguous records. We then report the distribution of the identified governance functions across the three projects and explain each function through representative records. Finally, we present a cross-cutting observation concerning the timing of public rule codification.

\subsection{Method}

We conducted a thematic analysis of the 265 project-level substantive rule-modifying records identified in RQ2 to examine the governance functions reflected in rule changes. For each record, we examined the semantic diff, the affected rule theme, and the surrounding context in the changed document. We use \textit{governance function} to denote the primary role performed by the modified rule-bearing text within project governance.

The coding proceeded in three rounds. First, the first author reviewed all records through open coding. The first and second authors then discussed the initial codes and developed a preliminary codebook containing category definitions, inclusion and exclusion criteria, boundary cases, and representative examples. Second, the first author recoded the complete dataset using this codebook, prepared analytical memos for each rule theme, and marked records with competing interpretations as \textit{ambiguous}. Of the 265 records, 24 were initially marked as ambiguous. Third, the first and second authors reviewed these records through multiple discussions, with unresolved cases referred to the third author. After refining the codebook, the first author re-examined all 265 records and produced the final coded dataset.

\subsection{Results}

Our analysis identified four recurring governance functions: \textit{Norm Alignment}, \textit{Workflow Refinement}, \textit{Coordination Structuring}, and \textit{Community and Governance Development}. Each record was assigned the function most directly reflected in its modified rule-bearing text. When a record could plausibly serve several functions, we selected the function expressed most centrally in the diff and document context.

As shown in \cref{tab:governance_functions}, Workflow Refinement was the most frequent function, accounting for 120 records (45.3\%). Community and Governance Development was particularly visible in Paddle, with 27 records, compared with 13 records in each of PyTorch and TensorFlow. PyTorch contained the largest number and proportion of Coordination Structuring records (18 records; 24.7\%). Norm Alignment accounted for 50 records overall, including 19 records in Paddle and 19 in TensorFlow.

\begin{table*}[t]
\centering
\caption{Distribution of governance functions across the three projects.}
\label{tab:governance_functions}
\begin{tabular}{lrrrr}
\toprule
\textbf{Governance function}
& \textbf{Paddle}
& \textbf{PyTorch}
& \textbf{TensorFlow}
& \textbf{Total} \\
\midrule
Norm Alignment
& 19 & 12 & 19 & 50 (18.9\%) \\

Workflow Refinement
& 29 & 30 & 61 & 120 (45.3\%) \\

Coordination Structuring
& 10 & 18 & 14 & 42 (15.8\%) \\

Community and Governance Development
& 27 & 13 & 13 & 53 (20.0\%) \\
\midrule
\textbf{Total}
& \textbf{85}
& \textbf{73}
& \textbf{107}
& \textbf{265 (100.0\%)} \\
\bottomrule
\end{tabular}
\end{table*}

\subsubsection{Norm Alignment}

Norm Alignment captures rule changes that codify widely established responses to baseline governance needs recurring across OSS projects. These needs arise from the basic conditions of open collaboration: participants need shared expectations for interaction, contributions need sufficient consistency and information to be reviewed, and community members need clear entry points for participation and communication. As OSS communities repeatedly encounter these needs, certain governance practices become commonly recognized ways of addressing them. We identified 50 records with this function.

Codes of Conduct illustrate this function in community interaction. For example, all three projects adopted the \textit{Contributor Covenant}, a Code of Conduct commonly used in open-source communities. The document established expectations for respectful and harassment-free participation and described how unacceptable behavior could be reported and handled. The relevant norm is the provision of explicit behavioral expectations for maintaining a safe and welcoming community environment.

Technical conventions address similarly recurring needs in development work. All three projects required contributors to follow coding styles. The governance function of this rule was to maintain consistency across code written by different contributors, thereby reducing the effort needed to understand and review contributions. The particular style guide may vary across projects, the underlying practice of establishing a common coding convention was shared across the three studied projects.

Basic contribution and communication mechanisms followed the same pattern. Issue and PR templates prompted contributors to provide a predictable minimum set of information needed to process submissions. Public mailing lists, forums, and issue trackers provided visible channels through which users and contributors could contact the community. Contributor License Agreements clarified the conditions under which external contributions could be accepted. Although their concrete implementations differed, these records addressed needs that commonly arise in OSS collaboration.

\subsubsection{Workflow Refinement}

Workflow Refinement refers to rule changes that improve how an established governance activity is carried out. These changes add, remove, or adjust concrete steps, exceptions, or execution instructions within a single activity, making the process clearer and more actionable while reducing ambiguity, rework, or repeated clarification. Its focus is on how an activity should be performed in practice.

We identified 120 records with this function, making it the most frequent governance function in all three projects. This concentration reflects the cumulative nature of documented workflows: after a contribution process is established, projects repeatedly add details to handle exceptions, clarify responsibilities within the activity, and reduce ambiguity in execution.

PyTorch's PR review process illustrates how a general workflow becomes progressively operationalized. Its early contribution guidance stated that contributors should iterate on a pull request until it was accepted and that the project team would merge it after approval and successful continuous-integration checks. Later records added rules for recurring situations within this process. For example, a reviewer who formally requested changes was expected to re-review the pull request within 24 hours after the author addressed the comments, because ``they are blocked on you.'' Another record instructed contributors to remind reviewers when a pull request had received no response for several business days, noting that area maintainers received many notifications and might otherwise overlook it. The project subsequently documented a fuller sequence covering triage, reviewer assignment, approval, trunk testing, post-merge failures, and reverts. Together, these changes transformed a general review expectation into a more explicit procedure that guided contributors and reviewers through common delays and exceptional situations.

Workflow Refinement also appeared in issue and PR handling. The studied projects expanded submission instructions and templates to request information such as the execution environment, reproduction procedure, and the relationship between a change and an existing issue. These additions gave maintainers more of the information needed to assess a submission and reduced the need for repeated clarification. Similar refinements appeared in code-acceptance criteria, documentation-writing requirements, issue-assignment procedures, and the maintenance of existing communication channels.

\subsubsection{Coordination Structuring}

Coordination Structuring refers to rule changes that organize relationships among multiple information flows, technical domains, participating roles, responsibility boundaries, or decision stages. These rules specify where different matters should be handled, who is responsible for particular areas, and how issues or proposals move across participants and stages. Their governance function is to make distributed collaboration manageable by establishing explicit routing, allocation, and coordination structures.

We identified 42 records with this function. These records commonly appeared when projects divided communication by purpose, formalized decision paths for substantial proposals, or distributed responsibility across technical modules and contributor groups. Instead of refining the execution of one activity, they structured how several activities, actors, or areas interacted.

TensorFlow's communication guidance illustrates the structuring of information flows. Shortly after the repository became public, the project separated help and support questions, general development discussions, and bug or feature reports. Support questions were directed to Stack Overflow, development discussions to a mailing list, and bugs or feature requests to the GitHub issue tracker. The document explicitly stated, ``Please do not use the mailing list or issue tracker for support.'' This rule established distinct destinations for different kinds of communication and clarified how each type of request should enter the project.

Formal feature-proposal processes provided a second form of coordination structure. TensorFlow's RFC process assigned a sponsor to guide a proposal and a review committee to recommend whether it should be adopted. It also connected proposal submission, community discussion, expert review, and final decision through a documented sequence. PyTorch and Paddle introduced comparable proposal mechanisms for substantial design changes. These rules created a distinct path for high-impact proposals and coordinated the roles of proposers, reviewers, and the wider community.

A third form involved the distribution of technical responsibility. TensorFlow and Paddle organized Special Interest Groups (SIGs) around particular technical or community areas, while PyTorch documented module maintainers. These structures assigned responsibility for reviewing changes, maintaining components, and coordinating work within defined domains. By dividing a large codebase or community into areas with identifiable ownership, the projects made responsibility and decision authority easier to locate.

\subsubsection{Community and Governance Development}

Community and Governance Development refers to rule changes that shape how people enter the community, continue participating, receive recognition, acquire formal roles, and take part in project governance. These rules create visible pathways between contribution, community membership, and governance authority. Their governance function is to sustain participation over time and to make the community's contributor structure and formal governance arrangements explicit. We identified 53 records with this function.

Task recommendation provided a direct entry point into participation. \textit{Good first issue} guidance identified small bug fixes and incomplete tasks as suitable starting points and described completing such work as a contributor's ``first step'' toward code contribution. The three projects also used documentation events, competitions, and other technical activities to make contribution opportunities visible. These rules connected potential contributors with concrete work and provided structured ways to begin participating in the project.

Recognition mechanisms made continued contribution visible within the community. Paddle documented that developers whose pull requests were merged would become Contributors and receive an open-source contribution certificate. Its \textit{Open Source Star} program recognized developers who had made substantial contributions during a given period. PyTorch similarly established a Contributor Award. These mechanisms linked participation with public recognition, contributor identity, and community status.

Community and Governance Development also covered the progression from contribution to formal responsibility. The projects documented qualifications for becoming committers, maintainers, or other roles, as well as nomination, application, and appointment procedures. These rules made contributor progression more visible by specifying how sustained participation could lead to greater responsibility.

\subsubsection{Observation: Public Codification May Lag Behind Governance Practice}

In addition to the four governance functions, we identified a cross-cutting temporal pattern: some document changes publicly codified governance arrangements that had already developed or were already being used in practice. In these cases, the commit gathered, articulated, or made publicly accessible expectations that had previously emerged through repeated collaboration, discussion, or established project routines. We refer to this gap as a \textit{lag in public codification}.

PyTorch's governance document provides one example. When it was added in 2019, it described an operating structure of project maintainers, core developers, and moderators, named the people occupying these roles, and documented their responsibilities and decision procedures. Although the document does not reveal when each arrangement first emerged, its present-tense description suggests that the commit consolidated an existing governance structure.

More explicit evidence appears in PyTorch's design-philosophy document, which described principles that had ``developed over time in PyTorch.'' TensorFlow showed a similar pattern. Its API-owner review document gathered ``commonly discussed topics'' from an existing twice-weekly review process, while its design-review criteria stated that the team had conducted internal and public reviews ``for a while.'' These records converted accumulated principles and recurring review experience into shared public references.

We treated a record as evidence of codification lag only when the added text indicated that the arrangement was already operating, had developed over time, or summarized recurring prior discussions. A late first appearance alone was not sufficient. This observation also clarifies how the first-appearance metric in RQ2 should be interpreted: it identifies the first appearance of a rule theme in our collected documentation, which marks its public codification but may postdate the emergence of the underlying practice.

\begin{findingbox}{Summary for RQ3}
Substantive rule changes reflected four recurring governance functions: Norm Alignment, Workflow Refinement, Coordination Structuring, and Community and Governance Development. Workflow Refinement was the most frequent, accounting for 120 of the 265 records. We also found that some governance arrangements and recurring review practices were publicly documented only after they had developed in practice. Accordingly, the first appearance captured in RQ2 marks public codification in the collected documents and may postdate the emergence of the underlying practice.
\end{findingbox}

\section{Actionable Governance Practices for OSS Communities}
\label{sec:suggestion}

Based on the integrated findings from RQ1--RQ3, we synthesize 33 actionable governance practices for OSS communities. The 17 rule themes identified in RQ1 specify the governance areas covered by these practices. The longitudinal findings from RQ2 inform when similar rules first appeared in documentation, were refined, or were reorganized in the studied projects. The four governance functions identified in RQ3 organize the practices according to the roles they perform in project governance: \textit{Norm Alignment}, \textit{Workflow Refinement}, \textit{Coordination Structuring}, and \textit{Community and Governance Development}. The complete list is provided in \cref{app:practices}.

The practices draw on three mature deep learning frameworks with complex contribution workflows, distributed collaboration, and substantial organizational participation. Their application depends on the technical and organizational conditions of an individual project.

\subsection{Establishing Shared Governance Baselines}

Norm Alignment practices establish the common expectations needed for public participation. They cover conduct, contribution procedures, communication, quality requirements, review, and repository authority.

Projects can publish expected standards of behavior and explain how violations are reported (P1). They can also document the basic contribution workflow (P2), provide visible communication channels (P3), and use templates to standardize issue and pull-request submissions (P4). Testing and review requirements make acceptance expectations visible (P5--P7), while repository role descriptions clarify who may review and accept contributions (P8).

\subsection{Refining Recurring Governance Workflows}

Workflow Refinement practices clarify how recurring contribution and maintenance activities are performed. They address changes in tools, task assignment, submission information, pull-request scope, quality requirements, review procedures, and role responsibilities.

Contribution instructions should be updated when the development or submission process changes (P9). Projects can specify how issues are claimed and reassigned (P10), and keep submission templates aligned with current review needs (P11). Communication entry points should also remain current (P12).

For code contributions, each pull request can be kept focused on one identifiable change (P13). Acceptance criteria and documentation style can then be refined around the technical and design concerns relevant to the project (P14--P15). Review procedures can likewise be updated as recurring cases emerge (P16). Projects can also define how feature proposals are assessed (P17) and revise role descriptions when operational responsibilities change (P18--P19).

\subsection{Structuring Cross-Role and Cross-Domain Coordination}

Coordination Structuring practices clarify how different requests, technical areas, and decision processes are connected.

Projects can direct different types of requests to appropriate channels (P20). Substantial design proposals can use dedicated templates and a documented review path (P21--P22). When technical ownership becomes distributed, projects can assign responsibility by module or area (P23). They can also establish SIGs or similar groups for persistent domains (P24), and define specific roles for multi-stage proposal or decision processes (P25).

\subsection{Developing Participation and Governance Structures}

Community and Governance Development practices define how contributors enter the project, receive support, gain recognition, and acquire formal responsibility.

Projects can provide entry-level tasks and structured contribution programs (P26--P27). Contributor-support roles can offer continuing guidance (P28). As participation becomes more formal, projects can publish role qualifications and acquisition procedures (P29--P30). Recognition mechanisms can make sustained participation visible (P31), while formal repository roles and governance bodies clarify how authority is distributed (P32--P33).

\section{Discussion}

Our study provides an empirical account of the content and evolution of documented governance rules and the governance functions reflected in their modification across three deep learning OSS projects. Beyond the actionable practices synthesized in \cref{sec:suggestion}, the findings have broader implications for project maintainers and researchers.

\subsection{Interactions Among Governance Rules}

Although the taxonomy distinguishes rule themes analytically, these rules do not operate independently. Some form complementary stages of the same workflow. For example, \textit{Task Recommendation} helps contributors identify suitable work, whereas \textit{Issue Assignment} coordinates commitment after a task has been selected. A project may therefore provide recommendations without preventing duplicated effort, or regulate task claiming without helping newcomers discover suitable tasks.

Similar interactions occur throughout the contribution process. \textit{Contribution Description Standards} specify what information contributors provide, \textit{Code Acceptance Criteria} define acceptable outcomes, and the \textit{Code Review Process} determines how those outcomes are assessed. Misalignment among these rules may create friction, for example when submission templates omit information required by reviewers or when submission guidance is inconsistent with acceptance criteria.

The documentation-splitting patterns observed in RQ2 further show that related rules are often reorganized around specific contributor workflows. Such specialization can improve task-local clarity, but it can also distribute interdependent requirements across multiple files, increasing the risk of inconsistency or reduced discoverability when one rule changes without corresponding updates elsewhere. Our study did not systematically model dependencies or conflicts among rule themes; examining these interactions is therefore an important direction for future work.

\subsection{Implications for Practitioners}

Our findings suggest that the development of governance rules is closely related to project context. Project scale, technical structure, contributor composition, organizational involvement, and development practices shape the coordination problems that become salient and the rules used to address them. Projects with relatively simple contribution workflows may rely primarily on visible submission guidance, communication channels, quality expectations, and clear review responsibilities. As contributor participation expands, technical responsibility becomes distributed, or work spans multiple modules and decision stages, projects may introduce more differentiated repository roles, module-level ownership, information-routing rules, and formal proposal procedures.

Technical and workflow changes also create recurring needs to revise governance rules. Changes in development tools, testing infrastructure, supported platforms, release processes, or documentation conventions may require maintainers to refine contribution, review, acceptance, and documentation requirements. Meanwhile, persistent contributor participation and more complex organizational structures may create a need for clearer role-acquisition procedures, contributor-support roles, recognition mechanisms, and formal governance bodies. The timing and form of these rules therefore vary with the coordination conditions faced by each project rather than following a uniform development sequence.

The organization of governance documentation should evolve alongside its rule content. Workflow- or role-specific documents can improve clarity by presenting rules in the context in which they are applied. However, when interdependent requirements are distributed across multiple files, contributors may have difficulty locating the complete workflow, and updates to one document may become inconsistent with related rules elsewhere. Projects can reduce these risks by providing clear navigation between related documents, identifying authoritative sources, and reviewing connected submission, acceptance, and review rules together when one of them changes.

\subsection{Implications for Researchers}

Firstly, our study highlights the value of moving beyond static descriptions of governance toward longitudinal analysis. Existing studies often characterize project governance using a single snapshot, whereas our findings show that different governance rules are publicly codified and revised at different stages of project development. This suggests that analyses based on one point in time may miss important changes in how projects organize contribution, coordination, and decision-making. Future research could therefore examine governance as an evolving process and investigate how the timing and sequence of rule codification relate to project conditions. For example, researchers could study whether delays in documenting particular rules are associated with recurring coordination difficulties, inconsistent contributor expectations, or greater maintenance effort. Such work would help clarify when changes in documented governance reflect routine refinement and when they signal broader shifts in project organization.

Secondly, the taxonomy of governance rules we constructed based on the IAD framework offers a theoretical basis for both comparative and explanatory research. By organizing documentary rules into seven rule types, the taxonomy provides a shared vocabulary for comparative research and supports the operationalization of governance attributes as measurable variables. Future studies could use these variables to examine associations between governance configurations and project outcomes. The research community can utilize this taxonomy to conduct cross-domain comparisons or causal modeling. For instance, future work could apply our classification scheme to contrast the governance evolution of deep learning frameworks against other types of software, such as operating systems or web libraries. Such comparative studies would help assess how broadly the identified evolutionary patterns apply across different OSS contexts.

Finally, our analysis identified cases in which public codification appeared to follow the emergence of governance practice. Some documents described arrangements that were already operating, principles that had developed over time, or recurring topics from existing review processes. This observation highlights a limitation of documentation-based governance research: repository histories can establish when a rule became publicly observable in the collected documents, but may not reveal when the underlying practice first emerged. Future research could investigate this gap by combining document histories with interviews, meeting records, mailing-list archives, and issue or pull-request discussions. Such evidence could help distinguish newly introduced rules from the later public articulation of established practices. Researchers could also examine whether mismatches between documented rules and actual project practices affect external contributors, for example by creating uncertainty about expected procedures or decision criteria.

\section{Threats to Validity}

\subsection{Construct Validity}
The primary threat is the potential misalignment between the IAD framework's seven rule types and the actual governance practices in OSS communities. To mitigate this, we did not force-fit data into pre-defined categories. Instead, we adopted an iterative coding strategy. We first conducted open coding to identify emerging themes from the raw documentation. Only after these themes were established did we map them to the IAD rule types. Throughout this process, we found that the identified governance themes could be mapped to the IAD framework without residual undefined categories, suggesting that the IAD taxonomy provided adequate coverage for the documentary governance rules observed in our dataset.

Another threat involves the completeness and accuracy of the rule evolution history (RQ2). Governance documentation in OSS projects often undergoes complex changes, such as file renaming, document reorganization, or migration across repositories and platforms, which may lead to broken history chains. To mitigate this, we did not rely solely on automated scripts. We applied a history reconstruction strategy that combined \texttt{git log --follow}, explicit migration clues such as redirection URLs and migration-related pull requests, and manual searches based on historical filenames, directory structures, and distinctive rule-bearing text fragments. When no predecessor path could be identified, we used the earliest available commit of the current file as the start of its traceable history. We also excluded trivial commits, such as typo fixes and formatting changes, to ensure that the identified modifications reflected substantive governance changes.

A related threat concerns the documentation-splitting metric. This metric captures the cumulative historical presence of rule themes across distinct logical files, rather than the exact number of active rule-bearing files in each year. It may therefore overestimate file distribution in rare cases where a document was deleted and its content later reappeared under a different name, or where a renamed document was partially split. To mitigate this threat, we normalized clear file renames and migrations as the same logical file and conservatively merged ambiguous migration cases when documents served the same contributor-facing function and showed substantial content overlap. We found that most documentation changes added or specialized rule-bearing content within existing document structures, while complete deletion-and-reappearance cases were uncommon. We therefore interpret documentation splitting as an indicator of historical documentation spread and specialization rather than a precise yearly count of active rule-bearing documents.

Another threat concerns the manual identification and interpretation of rule-modifying records. To reduce noise in RQ2, we applied explicit inclusion and exclusion criteria: we included records only when they substantively added, removed, or revised rule-bearing text, and excluded maintenance edits, formatting changes, document refactoring without substantive rule changes, and updates to concrete instances under existing rule mechanisms.

For RQ3, the main construct-validity concern lies in how the governance function of each rule change was operationalized. The modified text does not necessarily reveal the maintainers' complete intentions or the event that originally prompted the change. We therefore do not interpret the four governance functions as psychological motivations. Instead, they describe the primary governance role directly reflected in the modified rule-bearing text. Coding therefore relied primarily on semantic diffs, the affected rule theme, and the surrounding document context. In some cases, commit messages and linked issues or pull requests were consulted only as supplementary context; they were not treated as direct evidence of maintainers' complete intentions. Records supporting competing interpretations were initially marked as \textit{ambiguous} and resolved through author discussion using explicit category definitions, inclusion and exclusion criteria, boundary rules, and representative examples.

\subsection{Internal Validity}
The main internal validity concern arises from the interpretive nature of thematic coding in RQ1 and governance-function coding in RQ3. For RQ1, mapping heterogeneous rule-bearing text to the IAD framework could be influenced by researcher judgment. We reduced this risk through a structured coding procedure involving multiple authors, an explicit codebook, and discussion of disagreements. For RQ3, some records could plausibly reflect more than one governance function. We therefore used explicit category boundaries and representative examples, marked uncertain cases as \textit{ambiguous}, and resolved them through author discussion before rechecking the full dataset.

The second threat concerns the use of public documentation as evidence of actual governance practice. Our analysis identified cases in which documents described arrangements that were already operating, principles that had developed over time, or recurring topics from existing review processes. This suggests that a rule's first appearance in the collected documentation may postdate the emergence of the underlying practice. Accordingly, our analysis captures when governance rules became publicly codified in the collected documents, rather than the precise moment at which every practice first emerged. This limitation is particularly relevant to the first-appearance analysis in RQ2. At the same time, public documentation remains consequential because it constitutes the governance information directly available to external contributors. Our findings should therefore be interpreted as characterizing documented governance, while recognizing that undocumented internal practices may not be fully represented.

\subsection{External Validity}


Our empirical analysis is based on three mature deep learning frameworks: PyTorch, TensorFlow, and Paddle. These projects are large-scale, technically complex, rapidly evolving, and supported by substantial organizational or corporate participation. These characteristics make them information-rich cases for observing how governance rules are codified, organized, and revised under high coordination demands. At the same time, they also shape the transferability of our findings to other OSS contexts.

A further limitation concerns the temporal validity of our findings. Our dataset ends on December 31, 2024, and therefore reflects the documented governance of the three projects as of early 2025. Because some operational rules were modified frequently, subsequent changes to rule content, documentation structures, and evolution patterns are not captured. The findings should therefore be interpreted as a historical account up to the data-collection cutoff rather than as a description of the projects' current governance.

Some findings address governance needs that are common across many OSS projects. These include visible contribution entry points, issue and pull-request procedures, coding and testing expectations, communication channels, documentation organization, review workflows, and newcomer guidance. The IAD-based taxonomy may also support analysis beyond the studied domain because it classifies rules according to the institutional elements they regulate rather than project-specific technical content.

Other findings are more sensitive to project context. Practices concerning formal role progression, differentiated repository authority, module- or area-level ownership, SIGs or comparable subcommunities, multi-stage proposal processes, contributor recognition, and project-level governance bodies depend more strongly on project scale, contributor composition, technical modularity, coordination complexity, and organizational sponsorship. For example, a mid-sized community-led project may benefit from clear reviewer responsibilities and contribution standards, while formal subcommunities or multi-level repository roles may be less relevant to its current organization.

The technical domain also affects transferability. Deep learning frameworks often involve fast-moving dependencies, multiple hardware backends, performance requirements, release compatibility, and large downstream user bases. Governance practices shaped by these conditions may require adaptation when applied to projects in other domains, such as web frameworks, developer tools, infrastructure libraries, or small application projects. Future work could compare projects across different domains, sizes, maturity levels, and sponsorship models, including smaller projects, community-led projects, and projects with less centralized governance documentation.

\subsection{Validation of Actionable Practices}

The 33 actionable practices were synthesized from the rule themes identified in RQ1, the longitudinal patterns observed in RQ2, and the governance functions identified in RQ3. They therefore reflect recurring governance actions and conditions found in the three studied projects. Their practical relevance and effects, however, have not yet been examined through maintainer interviews, contributor surveys, independent expert evaluation, or field studies.

The need for further validation is especially important for practices whose application depends strongly on project context. These include the introduction of differentiated repository roles, module-level ownership, formal proposal procedures, specialized subcommunities, contributor-recognition mechanisms, and project-level governance bodies. Their relevance may vary with project scale, technical modularity, contributor composition, coordination complexity, and organizational sponsorship.

Future work could examine how maintainers assess, select, and adapt these practices across different OSS settings. Interviews and surveys could evaluate their perceived relevance and feasibility, expert review could identify missing or overly specific practices, and longitudinal field studies could investigate how individual practices affect contribution workflows, coordination, and governance maintenance over time.

\section{Conclusion}

In this paper, we examined how governance rules are codified, organized, and revised in three mature deep learning OSS projects: PyTorch, TensorFlow, and Paddle. To provide a structured account of documentary governance, we operationalized the IAD framework and combined thematic analysis of project documents with longitudinal analysis of their version histories.

For RQ1, we identified 17 rule themes across seven IAD rule types and examined how they were distributed across document categories and projects. The findings show that operational rule themes, such as contribution submission guidance and code acceptance criteria, were broadly shared and commonly documented in contributing files and templates. Structural and incentive-related rule themes, including role acquisition, governance authority, and recognition mechanisms, showed greater cross-project variation.

For RQ2, we reconstructed the histories of rule-bearing documents and analyzed substantive rule-modifying records. Operational rule themes generally appeared earlier in the collected documentation and were revised more frequently, whereas many structural rule themes appeared later and changed less often thereafter. We also observed increasing documentation specialization, as rule-bearing content became distributed across workflow-specific, role-specific, and subcommunity-specific documents.

For RQ3, we coded the primary governance function reflected in each substantive rule change. We identified four recurring functions: Norm Alignment, Workflow Refinement, Coordination Structuring, and Community and Governance Development. Workflow Refinement was the most frequent, reflecting repeated efforts to make contribution, review, testing, documentation, and maintenance procedures more explicit and executable. Coordination Structuring organized relationships among communication channels, technical domains, responsibilities, and decision stages, while Community and Governance Development shaped participation pathways, contributor recognition, formal roles, and governance institutions. We also identified cases in which public codification followed the development of governance practice, indicating that the first appearance in the collected documentation may not coincide with the origin of the underlying arrangement.

Based on the integrated findings from the three research questions, we synthesized 33 actionable governance practices for OSS maintainers. These practices connect the governance content identified in RQ1 with the public codification and revision patterns observed in RQ2 and the governance functions identified in RQ3. They are intended as an empirically grounded reference for projects facing coordination demands comparable to those of the three studied deep learning frameworks, rather than as universally applicable prescriptions. Future work could examine their transferability across other technical domains, project scales, and governance models, and validate their practical usefulness through maintainer interviews, surveys, expert review, or field studies.


\bibliographystyle{ACM-Reference-Format}
\bibliography{ref}

\appendix

\section{Complete List of Actionable Governance Practices}
\label{app:practices}

This appendix presents the 33 actionable governance practices synthesized from the findings of RQ1--RQ3. The practices are grouped by their primary governance function.

{
\footnotesize
\setlength{\tabcolsep}{4pt}
\renewcommand{\arraystretch}{1.08}
\setlength{\aboverulesep}{0pt}
\setlength{\belowrulesep}{0pt}

\begin{longtable}{
>{\centering\arraybackslash}p{0.7cm}
>{\raggedright\arraybackslash}p{3.3cm}
>{\raggedright\arraybackslash}p{9.0cm}}

\caption{Actionable governance practices for OSS communities.}
\label{tab:complete_practices} \\

\toprule
\textbf{ID} & \textbf{Rule Theme} & \textbf{Actionable Practice} \\
\midrule
\endfirsthead

\multicolumn{3}{c}{\small\textit{Table \thetable\ continued from previous page}} \\
\toprule
\textbf{ID} & \textbf{Rule Theme} & \textbf{Actionable Practice} \\
\midrule
\endhead

\bottomrule
\multicolumn{3}{r}{\small\textit{Continued on next page}} \\
\endfoot

\bottomrule
\endlastfoot

\multicolumn{3}{l}{\textbf{Norm Alignment}} \\
\midrule

P1
& Code of Conduct
& Publish a Code of Conduct that defines expected behavior, reporting channels, and procedures for handling reported violations. \\

P2
& Contribution Submission Guidance
& Document the baseline contribution workflow, including repository setup, pull-request submission, required agreements, and relevant development tools. \\

P3
& Communication Channels
& Provide visible public channels for project discussion, contributor questions, and issue reporting. \\

P4
& Contribution Description Standards
& Provide basic issue and pull-request templates that request a consistent minimum set of information. \\

P5
& Code Acceptance Criteria
& Publish baseline acceptance requirements concerning testing, coding style, compatibility, and continuous-integration checks. \\

P6
& Documentation Writing Style
& Adopt common documentation conventions, terminology, formatting rules, and external style guides where appropriate. \\

P7
& Code Review Process
& Document the basic review workflow, including reviewer selection, approval authority, and merging procedures. \\

P8
& Roles by Repository Write Access
& Identify the repository roles responsible for reviewing and accepting contributions, and document their permission boundaries. \\

\midrule
\multicolumn{3}{l}{\textbf{Workflow Refinement}} \\
\midrule

P9
& Contribution Submission Guidance
& Update contribution instructions when development tools, environments, CI procedures, or submission steps change. \\

P10
& Issue Assignment
& Specify whether issues may be claimed, how claims are recorded, and when inactive claims may be reassigned. \\

P11
& Contribution Description Standards
& Use issue and pull-request templates to collect the information needed for reproduction, evaluation, and review, and update their fields as project workflows and tooling change. \\

P12
& Communication Channels
& Keep channel links, contact details, and stated purposes current, and consolidate channels that are no longer maintained. \\

P13
& PR Size
& Keep each pull request focused on one identifiable change, and split large or unrelated changes into separate pull requests when possible. \\

P14
& Code Acceptance Criteria
& Refine acceptance criteria to cover project-specific concerns such as testing, supported environments, API behavior, backward compatibility, and design consistency. \\

P15
& Documentation Writing Style
& Specify project-specific requirements for documentation content, examples, terminology, build checks, and review. \\

P16
& Code Review Process
& Clarify reviewer assignment, response expectations, approval conditions, merging, reverts, and escalation procedures. \\

P17
& Feature Proposal Process
& Define how feature proposals are submitted and assessed, using a more formal review process for changes with broad architectural or compatibility implications. \\

P18
& Roles by Responsibility Scope
& Clarify the duties and operational limits of existing roles when their day-to-day responsibilities become unclear. \\

P19
& Roles by Repository Write Access
& Revise the responsibilities of repository roles when review, triage, approval, or merge procedures change. \\

\midrule
\multicolumn{3}{l}{\textbf{Coordination Structuring}} \\
\midrule

P20
& Communication Channels
& Route support questions, development discussions, bug reports, security reports, and substantial proposals to designated channels. \\

P21
& Contribution Description Standards
& Provide a dedicated proposal template when substantial design changes require information beyond an ordinary issue or pull request. \\

P22
& Feature Proposal Process
& Establish a documented review and decision path for substantial design changes. \\

P23
& Roles by Responsibility Scope
& Assign module- or area-level responsibility for review, maintenance, and technical decisions. \\

P24
& Roles by Responsibility Scope
& Establish SIGs or comparable subcommunities with explicit scopes and responsibility boundaries for persistent domains. \\

P25
& Roles by Responsibility Scope
& Define sponsor, committee, administrator, or similar roles for proposal and decision processes involving multiple stages. \\

\midrule
\multicolumn{3}{l}{\textbf{Community and Governance Development}} \\
\midrule

P26
& Task Recommendation
& Maintain visible entry-level tasks that provide newcomers with concrete contribution opportunities. \\

P27
& Task Recommendation
& Organize contribution campaigns, documentation events, internships, or similar programs around defined project work. \\

P28
& Roles for Contributor Support
& Establish maintained mentoring or contributor-support roles when contributors need continuing guidance. \\

P29
& Role Qualifications
& Publish the contribution, experience, conduct, and endorsement criteria used to evaluate candidates for formal roles. \\

P30
& Role Acquisition Process
& Document how formal roles are obtained, reviewed, renewed, or removed. \\

P31
& Incentive Mechanism
& Use awards or other recognition mechanisms to make sustained contribution visible within the community. \\

P32
& Roles by Repository Write Access
& Differentiate repository roles such as triager, committer, approver, and maintainer, and specify their associated permissions. \\

P33
& Roles by Responsibility Scope
& Publish the composition, responsibilities, authority, and decision scope of project-level governance bodies. \\

\end{longtable}
}

\end{document}